\documentclass[10pt,aps,fleqn,superscriptaddress,notitlepage,nofootinbib,preprintnumbers,nobalancelastpage]{revtex4-1}
\pdfoutput=1
\usepackage{amsmath,amssymb,pstricks,graphicx,xspace,subfigure}
\usepackage{todonotes}
\usepackage{bigints}

\newcommand{\mc}[1]{\mathcal{#1}}
\newcommand{\mr}[1]{\mathrm{#1}}

\newcommand{\Caesar}{C\scalebox{0.85}{AESAR}\xspace}
\usepackage{hyperref}
\hypersetup{
  pdfauthor={Stefan Hoeche, Daniel Reichelt, Frank Siegert},
  pdftitle={Momentum conservation and unitarity in parton showers and NLL resummation},
  pdfkeywords={QCD, Resummation, Parton Shower}
}
\begin{document}
\preprint{SLAC-PUB-17173}
\preprint{MCNET-17-20}
\title{Momentum conservation and unitarity in parton showers and NLL resummation}
\author{Stefan~H{\"o}che}
\affiliation{SLAC National Accelerator Laboratory,
  Menlo Park, CA, 94025, USA}
\author{Daniel~Reichelt}
\affiliation{Institut f{\"u}r Kern- und Teilchenphysik,
  01069 Dresden, Germany}
\author{Frank Siegert}
\affiliation{Institut f{\"u}r Kern- und Teilchenphysik,
  01069 Dresden, Germany}
\begin{abstract}
  We present a systematic study of differences between NLL resummation and
  parton showers. We first construct a Markovian Monte-Carlo algorithm
  for resummation of additive observables in electron-positron annihilation.
  Approximations intrinsic to the pure NLL result are then removed, in order
  to obtain a traditional, momentum and probability conserving parton shower
  based on the coherent branching formalism.
  The impact of each approximation is studied, and an overall comparison is made
  between the parton shower and pure NLL resummation. Differences compared
  to modern parton-shower algorithms formulated in terms of color dipoles
  are analyzed. 
\end{abstract}
\maketitle

\section{Introduction}
Searches for new physics and measurements of Standard Model parameters at the
Large Hadron Collider and possible future colliders require ever increasing precision
in the analysis of multi-scale events. Large scale hierarchies in such reactions will
generally result in large Bremsstrahlung effects. In order to reliably predict
measurable quantities, such as a fiducial cross section, the radiative corrections
determined in QCD perturbation theory must be resummed to all orders. Resummation
was first performed for energy-energy correlations in $e^+e^-$ collisions~\cite{
  Dokshitzer:1978hw,Parisi:1979se,Collins:1981uk}, transverse momentum
dependent cross sections in Drell-Yan events~\cite{Sterman:1986aj,Catani:1989ne,Catani:1990rp}
and $e^+e^-$ hadronic event shapes~\cite{Catani:1992ua}. Several observables
in hadron collisions have also been resummed analytically~\cite{Catani:2000jh}.
Such calculations have been extended to very high precision and used, for example,
to extract the strong coupling from experimental data in
$e^+e^-$ annihilation to hadrons~\cite{Dissertori:2009ik,Gehrmann:2009eh}.
Effective field theory methods~\cite{Bauer:2001yt,Bauer:2002nz} also contribute
to rapid progress in this field. General semi-analytic approaches to the problem
have been constructed~\cite{Kidonakis:1997gm,Kidonakis:1998nf,Laenen:1998qw,
  Bonciani:2003nt,Banfi:2001bz,Banfi:2004yd,Banfi:2014sua} and automated~\cite{Gerwick:2014gya}
based on direct QCD resummation. They depend only on universal coefficients
and are applicable to different processes and a large class of observables.
An alternative to analytical calculations is the simulation
of events in a Markov-Chain Monte-Carlo known as a parton shower\cite{
  Fox:1979ag,Marchesini:1983bm,Sjostrand:1985xi,Marchesini:1987cf}.
While the formal precision of this approach is comparable to analytic resummation
only in processes with a trivial color structure at the leading order, 
parton-showers typically give a good description of experimental measurements
and are therefore an integral part of the high-energy physics toolkit.

Even in the simplest scenarios the resummation performed by a parton shower
is not identical to an analytic computation. This study will investigate
the differences in some detail. We first show how a parton shower can be constructed
that reproduces the pure next-to-leading logarithmic (NLL) resummed result
as obtained by the semi-analytic \Caesar formalism \cite{Banfi:2004yd}.
For simplicity we will focus on additive observables in $e^+e^-$ annihilation to jets.
Starting from this algorithm, we successively include effects beyond NLL accuracy
that arise from momentum and probability conservation, such that a traditional
parton shower in the coherent branching formalism is recovered eventually.
To our knowledge this is the first time that a systematic study of this type 
has been performed. While we focus on a very simple setup, for which
parton showers have been shown to achieve NLL accuracy~\cite{Catani:1990rr},
we argue that most differences investigated here will also arise in more
complicated scenarios, such as hadron-hadron collisions and processes
with a non-trivial color structure at the Born level.
They will impact any prediction made for the Large Hadron Collider
and possible future colliders, and -- while formally sub-leading --
they may be numerically large and should be taken into account
as a systematic uncertainty.

This paper is organized as follows: Section~\ref{sec:comparison} recalls
those parts of the \Caesar formalism and of the parton shower formalism needed 
in this study. Section~\ref{sec:cross-check} presents the technical details
of a modified parton shower reproducing exactly the analytic NLL result.
Section~\ref{sec:analysis} analyzes the role of NLL approximations in detail
by removing them from the previously constructed shower one-by-one.
Section~\ref{sec:DGLAPvsCS} compares the full parton-shower result against
a more conventional parton-shower implementation, where soft double counting
is removed by partial fractioning of the soft eikonal.
Section~\ref{sec:conclusions} presents our conclusions.

\section{NLL resummation and the parton shower formalism}
\label{sec:comparison}
We first review the methods used for analytic resummation in \Caesar \cite{Banfi:2004yd}
as well as the parton shower algorithm \cite{Fox:1979ag,Marchesini:1983bm,
  Sjostrand:1985xi,Marchesini:1987cf}. They are cast into a common language
in order to allow an easy comparison between the two. We focus on the
simplest case of resummation of a 2-jet observable in $e^+e^-\to \mathrm{jets}$,
i.e.\ resummation of soft gluons emitted from a pair of two hard quark lines.

\subsection{Prerequisites and notation}
Following the \Caesar formalism, we denote the momenta of the hard partons as
$p_1, \ldots, p_n$. Additional soft emissions are denoted by $k$, and the observable
we wish to compute by $v$. In general, the observable will be a function of both
the hard and the soft momenta, $v=V(\{p\},\{k\})$, while in the soft approximation it
reduces to a function of the soft momenta alone, $v=V(\{k\})$. In the rest frame of
two hard legs, $i$ and $j$, considered to be the radiating color dipole, we can 
parametrize the momentum of a single emission as
\begin{equation}\label{eq:sudakov_decomposition}
  k=z_{i,j} p_i+z_{j,i} p_j+k_{T,ij}\;,
  \qquad\text{where}\qquad
  k_{T,ij}^2=2p_ip_j\,z_{i,j}\,z_{j,i}\;.
\end{equation}
We define the rapidity of the emission in this frame as $\eta_{ij}=1/2\ln(z_{i,j}/z_{j,i})$.
The observable, computed as a function of $k$ when radiated collinear
to the hard parton, $l$, can then be written as\footnote{%
  Note that because of the simplified setup that we use for this comparison,
  the dependence on $d_lg_l(\phi^{(l)})$ has been dropped, and that
  we will use $b=b_l$ in the following.}.
\begin{equation}\label{eq:V_approx}
  V(k) = \left(\frac{k_{T,l}}{Q}\right)^a e^{-b_l\eta_l}\;,
\end{equation}
where, in the collinear limit, we have $k_{T,l}=k_{T,lj}$ and $\eta_l=\eta_{lj}$
for any $j\nparallel l$. We restrict our analysis to the case of additive observables,
which can be calculated in the presence of multiple soft gluons as a simple sum,
$V\left(k_1,\dots,k_n\right) = \sum_i^n V(k_i)$. Such observables are
of great interest phenomenologically and, while relatively easy to compute,
already exhibit most complications associated with the effects of NLL approximations.

The parton shower used in our study will be based on DGLAP evolution 
\cite{Gribov:1972ri,Lipatov:1974qm,Dokshitzer:1977sg,Altarelli:1977zs}.
At NLL, for recursive infrared and collinear safe observables, gluon splitting
only contributes at the inclusive level and is therefore taken into account
effectively by working in the CMW scheme~\cite{Catani:1990rr}. In analogy to the
NLL \Caesar formalism, our parton shower will therefore only implement
gluon radiation off the hard partons, and soft double counting will be removed
by sectorization of the soft-emission phase space. Technical details are given
in Sec.~\ref{sec:implementation}, and a comparison to more conventional parton showers,
which include gluon splitting, is performed in Sec.~\ref{sec:DGLAPvsCS}.
The basis for DGLAP evolution are the collinear factorization properties of QCD
matrix elements. With $|\mc{M}_n(1,\ldots,n)|^2$ being the squared $n$-parton
matrix element, the factorization formula in the limit that partons $i$ and $j$
become collinear reads
\begin{equation}
  \mathop{d\Phi_{n+1}}\left|\mathcal{M}_{n+1}(1,..,i,..,j,..,n)\right|^2 \approx
  \mathop{d\Phi_{n}}\left|\mathcal{M}_{n}(1,..,ij,..,n)\right|^2\;
  \frac{\mathop{dt}}{t}\mathop{dz}\frac{\mathop{d\phi}}{2\pi}\,
  \frac{\alpha_s}{2\pi}P_{ij\,i}(z)\;.
\end{equation}
In this context, $\mathop{d\Phi_{n}}$ is the $n$-particle phase space element,
and $P_{ij\,i}(z)$ is the Altarelli-Parisi splitting kernel associated with the
branching of an intermediate parton $ij$ into partons $i$ and $j$.
Except for the analysis in Sec.~\ref{sec:DGLAPvsCS}, the only relevant
splitting kernel in our study is the quark-to-quark transition
\begin{equation}
  P_{qq}\left(z\right) = C_F\left[\frac{2}{1-z}-\left(1+z\right)\right]~.
\end{equation}
The treatment of gluon radiators is discussed in App.~\ref{sec:gluon_radiators}.
We denote the unregularized splitting probability between two scales, $t$ and $t'$, as
\begin{equation}
  R(t',t) = \int_{t'}^t \frac{\mathop{d\bar{t}}}{\bar{t}}\; R'(\bar{t})\;
  \qquad\text{where}\qquad
  R'(t) = \int_{z_{\rm min}(t)}^{z_{\rm max}(t)} dz\; \frac{\alpha_s}{2\pi}P_{qq}(z)\;.
\end{equation}
Following standard practice to improve the logarithmic accuracy of the resummation,
the strong coupling is evaluated at the transverse momentum of the gluon~\cite{Amati:1980ch},
and the soft enhanced term of the splitting functions is rescaled by $1+\alpha_s/(2\pi) K$,
where $K=(67/18-\pi^2/6)\,C_A-10/9\,T_R\,n_f$~\cite{Catani:1990rr}. The latter method
is known as the CMW scheme.

The integration boundaries for $z$ depend on the evolution variable and are given
by the constraint that the momentum in the anti-collinear direction must be preserved.
For the case of evolution in collinear transverse momentum, $k_T^2=2p_ip_j\,z(1-z)$,
we obtain $z_{\rm min/max}=(1\mp\sqrt{1-4k_T^2/Q^2})/2$ (cf.\ Sec.~\ref{sec:analysis}).
The probability for no splitting between two scales can be inferred from a unitarity
constraint, i.e.\ the condition that the parton shower be probability conserving.
For final-state evolution the no-branching probability is given by
\begin{equation}\label{eq:sudakov}
  \Pi(t',t) = e^{-R(t,t')}\;.
\end{equation}
Note that this particular form of the no-branching probability is equivalent
to the Sudakov form factor only at leading order, cf.\ App.~\ref{sec:gluon_radiators}.
Since we neglect gluon splitting, the functional form of $R$ is unchanged 
until the shower terminates, which greatly simplifies the calculation%
\footnote{In the general case of multiple hard legs the situation is
  complicated by the need to perform non-abelian exponentiation of next-to-leading
  logarithmic corrections originating in soft-gluon interference~\cite{
    Botts:1989kf,Kidonakis:1996aq,Kidonakis:1998bk,Oderda:1999kr,Kidonakis:2000gi}}.
The parton shower algorithm solves for the scale $t'$, based on a starting
scale $t$ and the total branching probability (differential in $\ln t$), 
\begin{equation}\label{eq:PSprob}
  \mc{P}(t',t) = \frac{d\,\Pi(t',t)}{d\ln t'}\;.
\end{equation}
It terminates when a cutoff scale $t_c$ is reached. Typically, $t_c$ is defined
such as to mark the transition to the non-perturbative regime, i.e.\ the region
where $\alpha_s/(2\pi)\approx 1$.

\subsection{Casting analytic resummation into the parton shower language}
To enable a comparison with the semi-analytic resummation framework of \Caesar,
we consider the cumulative cross section in an arbitrary observable, $v$, defined as
\begin{equation}\label{eq:cumulative_xs}
  \Sigma\left(v\right) := \frac{1}{\sigma}\int^v \mathop{d\bar{v}} 
  \frac{\mathop{d\sigma}}{\mathop{d\bar{v}}}\;.
\end{equation}
The calculation is simplified by choosing a parton shower evolution variable, $\xi$, 
that (up to a power) corresponds to $V(k)$
\begin{equation}
  \xi=k_T^2\,(1-z)^{-\frac{2b}{a+b}}\;.
\end{equation}
This implies that splittings giving the largest contribution to the observable are produced first.
Note that here and in the following we use $k_T=k_{T,l}$ and $\eta=\eta_l$.

If the effects of multiple emissions could be ignored, the cumulative cross section in
Eq.~\eqref{eq:cumulative_xs} would be given by the square of the survival probability,
Eq.~\eqref{eq:sudakov}, corresponding to the fact that radiation of a single gluon can
originate from either of the two hard legs in the two-quark leading-order final state.
It would then be sufficient to compute the probability $R(v)=R(v,1)$ for emissions resulting
in observable values larger than $v$. Already at the level of a single emission this would
lead to double counting~\cite{Marchesini:1983bm}. The problem can be circumvented
by sectorizing the phase space using the requirement $\eta>0$.
Note that this constraint is not strictly necessary for the collinear part of
the splitting function if the parton shower implementation is capable of handling
negative weights. However, this is not the case for most traditional shower algorithms,
which prompts us to apply the condition to the entire splitting function.
The combined probability for a single emission from any of the two hard legs
at $\xi>Q^2v^{2/(a+b)}$ can then be written as
\begin{equation}\label{eq:single_emission_ps}
  R_{\rm PS}(v) = 2\int_{Q^2v^\frac{2}{a+b}}^{Q^2}
  \frac{\mathop{d\xi}}{\xi}\; \int_{z_{\rm min}}^{z_{\rm max}} dz\;
  \frac{\alpha_s\big(\xi(1-z)^\frac{2b}{a+b}\big)}{2\pi}\,
  C_F\left[\frac{2}{1-z}-(1+z)\right]\,
  \Theta\bigg(\!\ln\frac{(1-z)^\frac{2a}{a+b}}{\xi/Q^2}\bigg)\;.
\end{equation}
This should be compared to Eq.~(2.17) of Ref.~\cite{Banfi:2004yd},
which can be rewritten in our parametrization as
\begin{equation}\label{eq:single_emission_nll}
  R_{\rm NLL}(v) = 2\int_{Q^2v^\frac{2}{a+b}}^{Q^2} \frac{\mathop{d\xi}}{\xi}\; 
  \left[\,\int_0^1 dz\;
    \frac{\alpha_s\big(\xi(1-z)^\frac{2b}{a+b}\big)}{2\pi}\frac{2\,C_F}{1-z}
    \Theta\bigg(\!\ln\frac{(1-z)^\frac{2a}{a+b}}{\xi/Q^2}\bigg)
    -\frac{\alpha_s(\xi)}{\pi}\,C_FB_q\right]\;.
\end{equation}
A brief summary of semi-analytic resummation based on~\cite{Banfi:2004yd} and
using Eq.~\eqref{eq:single_emission_nll} can be found in App.~\ref{sec:caesar}.
The no-emission probability based on $R_{\rm NLL}(v)$ can also be computed 
in a Markovian Monte-Carlo simulation, by starting from the parton-shower expression, 
Eq.~\eqref{eq:single_emission_ps}, and performing the following manipulations:
\begin{itemize}
\item The $z$-integration in the soft term runs from 0 to $1-(\xi/Q^2)^{(a+b)/2a}$,
  where the upper bound stems from the requirement that $\eta>0$ 
  (the $\Theta$-function in Eq.~\eqref{eq:single_emission_ps}), eliminating
  the double counting of soft-gluon radiation. 
\item The collinear term proportional to $(1+z)$ is integrated from 0 to 1
  in order to produce the collinear anomalous dimension, $B_q$. At the same time,
  $\alpha_s$ is evaluated at $\xi$.
\end{itemize}
Note in particular that the $z$-integration is extended beyond the values
$z_{\rm min}$ (and $z_{\rm max}$ in the collinear case) allowed by local
four-momentum conservation. This will be one of the effects investigated
in Sec.~\ref{sec:analysis}. 

The complete parton-shower prediction of the cumulative cross section, $\Sigma(v)$,
including effects from arbitrarily many emissions, and using the approximation
$V(\{p\},\{k\})=\sum_i V(k_i)$ is given by
\begin{equation}\label{eq:SigmaShower}
  \begin{aligned}
    \Sigma_\mathrm{PS}\left(v\right) &=
    \sum_{m=0}^{\infty}
    \left(\prod_{i=1}^{m}
      \int_{\xi_c}^{\xi_{i-1}}\frac{\mathop{d\xi_i}}{\xi_i}\, R_{\rm PS}'(\xi_i)\,e^{-R_{\rm PS}(\xi_{i-1},\xi_i)}
      \right)e^{-R_{\rm PS}(\xi_m,\xi_c)}\;\Theta\bigg(v-\sum_{j=1}^m V(t_j)\bigg)\bigg|_{\xi_0=Q^2}\\ 
    &=   e^{-R_{\rm PS}(Q^2,t_c)}  \sum_{m=0}^{\infty}
    \frac{1}{m!}\left(\prod_{i=1}^{m}
    \int_{t_c}^{Q^2}\frac{\mathop{dt_i}}{t_i}\, R_{\rm PS}'(t_i)\right)
    \Theta\bigg(v-\sum_{j=1}^m V(t_j)\bigg)\;.
  \end{aligned}
\end{equation}
We compare Eq.~\eqref{eq:SigmaShower} to the main result of \cite{Banfi:2001bz},
which reads
\begin{equation}\label{eq:SigmaCaesar}
  \Sigma_\mathrm{NLL}\left(v\right) = e^{-R_{\rm NLL}(v)}\mathcal{F}\left(v\right)\;.
\end{equation}
The exponential corresponds to the pure survival probability in terms of
Eq.~\eqref{eq:single_emission_nll}. The function $\mathcal{F}\left(v\right)$ 
accounts for the effect of multiple emissions. For the simple observables
considered here it can be written as \cite{Banfi:2004yd}\footnote{
  The $\epsilon \to 0$ limit can be taken analytically~\cite{Banfi:2001bz},
  cf.\ App.~\ref{sec:caesar}, Eq.~\eqref{eq:f_additive}.}:
\begin{equation}\label{eq:f_function}
  \mathcal{F}\left(v\right) = \lim_{\epsilon\to 0}
  \mathcal{F}_\epsilon \left(v\right)\;,
  \qquad\text{where}\qquad
  \mathcal{F}_\epsilon \left(v\right)= e^{R_{\rm NLL}'(v)\ln\epsilon}\sum_{m=0}^{\infty} \frac{1}{m!}
  \left(\prod_{i=1}^m R_{\rm NLL}'(v) \int_\epsilon^1 \frac{\mathop{d\zeta_i}}{\zeta_i}\right)
  \Theta\bigg(1-\sum_{j=1}^m \zeta_j\bigg)\;.
\end{equation}
Following the notation of Ref.~\cite{Banfi:2004yd}, $R_{\rm NLL}'(v)$ is the
derivative of $R$ with respect to $L=-\ln v$, excluding all terms formally 
not relevant at NLL accuracy. Note that Eq.~\eqref{eq:f_function} is a pure NLL
contribution to $\Sigma_{\rm NLL}(v)$, as $R'(v)$ by itself is sub-leading.
If we intend to generate  Eq.~\eqref{eq:f_function} using a parton shower,
the branching probability, Eq.~\eqref{eq:single_emission_ps}, must be modified
such as to reflect the differentiation w.r.t.\ the lower integration
limit in Eq.~\eqref{eq:single_emission_nll}, which leads to $\xi=Q^2v^{2/(a+b)}$,
as well as the condition that higher logarithmic terms are dropped in $R'(v)$.
We can satisfy these constraints using the following modifications of the plain
parton shower:
\begin{itemize}
\item The $z$-integration in the soft term runs from 0 to $1-v^{1/a}$.
\item The strong coupling runs at one loop and is evaluated at $v^{2/(a+b)}(1-z)^{2b/(a+b)}$.
\item The collinear term is dropped.
\end{itemize}
We can now rewrite Eq.~\eqref{eq:SigmaCaesar} in a form that is similar to
Eq.~\eqref{eq:SigmaShower}
\begin{equation}\label{eq:SigmaGeneral}
        \Sigma\left(v\right) = \exp\left\{-\int_v\frac{\mathop{d\xi}}{\xi} R'_{>v}(\xi)
          - \int_{v_\mathrm{min}}^v  \frac{\mathop{d\xi}}{\xi}
          R'_{<v}(\xi)\right\} \times\sum_{m=0}^\infty \frac{1}{m!}\left(\prod_{i=1}^{m}\int_{v_\mathrm{min}}
        \frac{\mathop{d\xi_i}}{\xi_i} R'_{<v}(\xi_i)\right)\Theta\bigg(v-
          \sum_{j=1}^m V(\xi_j)\bigg)\;.
\end{equation}
with $R'$ given by
\begin{equation}
        R'_{\lessgtr v}(\xi) =
        \frac{\alpha_s^{\lessgtr v,\mathrm{soft}}\big(\mu^2_\lessgtr\big)}{\pi}
        \int_{z^{\rm min}}^{z^\mathrm{max}_{\lessgtr v,\mathrm{soft}}} \mathop{dz} \frac{C_\mathrm{F}}{1-z}
        -  \frac{\alpha_s^{\lessgtr v,\mathrm{coll}}\big(\mu^2_{\lessgtr v}\big)}{\pi}
        \int_{z^{\rm min}}^{z^\mathrm{max}_{\lessgtr v,\mathrm{coll}}} \mathop{dz} C_\mathrm{F}\frac{1+z}{2}\;.
\end{equation}
\begin{table}[t]
\centering
\begin{tabular}{c|@{\quad}c@{\qquad}c@{\qquad}cc@{\qquad\qquad}c|@{\quad}c@{\qquad}c@{\qquad}c}
  &Resummation&Parton Shower&Figure& & &Resummation&Parton Shower&Figure\\
  \cline{1-4}\cline{6-9}
  $z^\mathrm{max}_{>v,\mathrm{soft}}$&\multicolumn{2}{c}{$1-(\xi/Q^2)^\frac{a+b}{2a}$}&n.a.
  && $z^\mathrm{max}_{>v,\mathrm{coll}}$&$1$&$1-(\xi/Q^2)^\frac{a+b}{2a}$&\ref{fig:kinConstSoftFinite}\\
  $\mu^2_{>v,\mathrm{soft}}$&\multicolumn{2}{c}{$\xi(1-z)^\frac{2b}{a+b}$}&n.a.
  && $\mu^2_{>v,\mathrm{coll}}$&$\xi$ &$\xi(1-z)^\frac{2b}{a+b}$ &\ref{fig:kinConstSoftFinite}\\
  $\alpha_s^{>v,\mathrm{soft}}$&\multicolumn{2}{c}{2-loop CMW}&n.a.
  && $\alpha_s^{>v, \mathrm{coll}}$&1-loop&2-loop CMW & \ref{fig:AlphaS}\\
  \cline{1-4}\cline{6-9}
  $z^\mathrm{max}_{<v,\mathrm{soft}}$&$1-v^\frac{1}{a}$&$1-(\xi/Q^2)^\frac{a+b}{2a}$&\ref{fig:kinConstSoftEnhanced}
  && $z^\mathrm{max}_{<v,\mathrm{coll}}$&$0$&$1-(\xi/Q^2)^\frac{a+b}{2a}$& \ref{fig:FullResult}\\
  $\mu^2_{<v,\mathrm{soft}}$&$Q^2v^\frac{2}{a+b}(1-z)^\frac{2b}{a+b}$&$\xi(1-z)^\frac{2b}{a+b}$&\ref{fig:kinConstSoftEnhanced}
  && $\mu^2_{<v,\mathrm{coll}}$&n.a. &$\xi(1-z)^\frac{2b}{a+b}$& \ref{fig:FullResult} \\ 
  $\alpha_s^{<v,\mathrm{soft}}$&1-loop&2-loop CMW& \ref{fig:AlphaS}
  && $\alpha_s^{<v, \mathrm{coll}}$&n.a.&2-loop CMW& \ref{fig:FullResult}\\
\end{tabular}
\caption{Choices of parameters in Eq.~\eqref{eq:SigmaGeneral} leading to
  Eq.~\eqref{eq:SigmaCaesar} (NLL resummation) and Eq.~\eqref{eq:SigmaShower} (parton shower).
  The effects of switching between the two parametrizations are investigated
  in the figure referred to in the last column. More details can be found in
  Sec.~\ref{sec:analysis}.}\label{tab:Parameters}
\end{table}
The choices of $\alpha_s$, $z^{\rm max}$ and $\mu^2$ corresponding to NLL resummation
in the \Caesar formalism and in a DGLAP-based parton shower are given in Tab.~\ref{tab:Parameters}.
The physical limits on the $z$-integral in Eq.~\eqref{eq:single_emission_ps}, which are a consequence
of local four-momentum conservation, are not easily formulated in terms of $\xi$ and will be investigated
separately in Sec.~\ref{sec:analysis}. It is interesting to note that $\mathcal{F}\left(v\right)$
by itself can be extracted from the same formalism by starting the shower evolution at $Q^2v^{\frac{2}{a+b}}$.
This fact has been used in the past to construct a dipole shower for the resummation of
non-global logarithms~\cite{Dasgupta:2001sh}.

\section{Markov-Chain Monte Carlo implementation}\label{sec:implementation}
\label{sec:cross-check}
As described in Sec.~\ref{sec:comparison}, the NLL resummation is nearly equivalent 
to a parton shower at the single-emission level. The differences lie in the treatment 
of the collinear term and of the lower integration boundary on $z$. These differences
also introduce a change in the scale of the running coupling
in Eq.~\eqref{eq:SigmaGeneral}. The choice of integration boundaries in the analytic
resummation implies that the splitting function turns negative in parts of the 
phase space. To deal with this situation in the Monte Carlo simulation, 
we use the methods discussed in \cite{Hoeche:2009xc,Lonnblad:2012hz}.
Splittings are generated according to an overestimate of the strong coupling
and the splitting kernel
\begin{equation}
  \alpha_s^{\rm max}P_\mathrm{max}\left(z\right) = \alpha_s^{\rm max}C_\mathrm{F}
  \left[\frac{2}{1-z}\Theta\left(z_\mathrm{max}'-z\right)+\gamma\Theta\left(z-z_\mathrm{max}'\right)\right]
\end{equation}
with an in principle arbitrary constant $\gamma$. For practical calculations we choose
$\gamma=2$. Note that the values of $z^\mr{max}_\mr{soft}$ and
$z_\mr{coll}^\mr{max}$ are overestimated by a common value in $P_\mathrm{max}$,
which we have made explicit by writing $z_{\rm max}'$. Splittings are vetoed
with a constant probability $1/C$ and are associated with a weight 
\begin{equation}\label{eq:ps_weight_corr}
  \omega = \frac{C\; \alpha_s^{\rm res} P_\mathrm{res}\left(z\right)}{
    \alpha_s^{\rm max}P_\mathrm{max}\left(z\right)}\times\left\{\begin{array}{cc}
  \displaystyle1 & \text{if accepted}\\[2mm]
  \displaystyle\frac{\alpha_s^{\rm max}P_\mathrm{max}\left(z\right)-\alpha_s^{\rm res}P_\mathrm{res}\left(z\right)}{
    \left(C-1\right)\alpha_s^{\rm res}P_\mathrm{res}\left(z\right)} & \text{if rejected}
  \end{array}\right.
\end{equation}
This correction accounts in particular for the negative sign of the integrand,
Eq.~\eqref{eq:P_res}, in the region $z>z_{\rm max}^{\rm soft}$. In addition, it
is possible to veto emissions violating the condition $\sum_i V(k_i) < v$, which
would contribute with zero weight, to improve numerical accuracy
\cite{Lonnblad:2012hz}. The value of $C$ determines how many emissions are
proposed, and thus potentially vetoed. It can again in principle be an arbitrary
constant larger than $1$, but is relevant for the speed of convergence. We choose
$C=2$ in our implementation.\\
The kernel eventually used for NLL resummation is given by
\begin{equation}\label{eq:P_res}
  \alpha_s^{\rm res}P_\mathrm{res} = C_\mathrm{F}
  \left[\alpha_s(\mu^2_{\rm soft})\frac{2}{1-z}\Theta\left(z^\mathrm{max}_{\rm soft}-z\right)
    -\alpha_s(\mu^2_{\rm coll})(1+z)\Theta\left(z^\mathrm{max}_{\rm coll}-z\right)\right]\;,
\end{equation} 
with $z_\mathrm{max}$ and $\mu^2$ chosen according to Table \ref{tab:Parameters}. 

\begin{figure}[t]
  \subfigure[~Thrust]{\begin{minipage}{0.32\textwidth}
    \includegraphics[width=\textwidth]{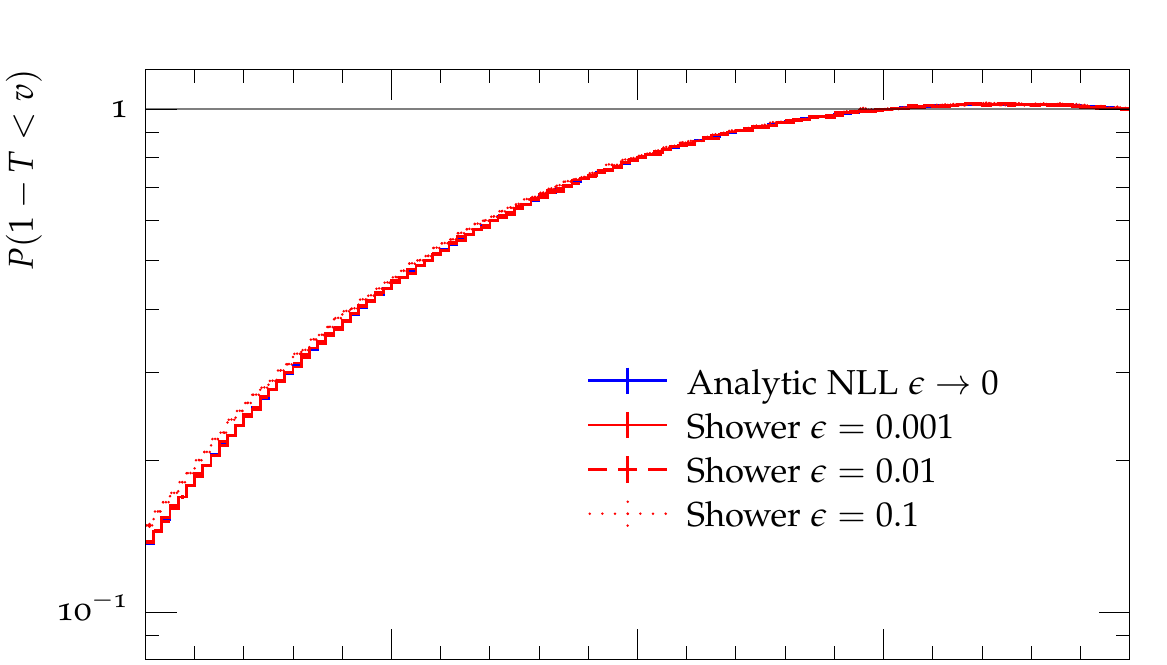}\\[-0.77mm]
    \includegraphics[width=\textwidth]{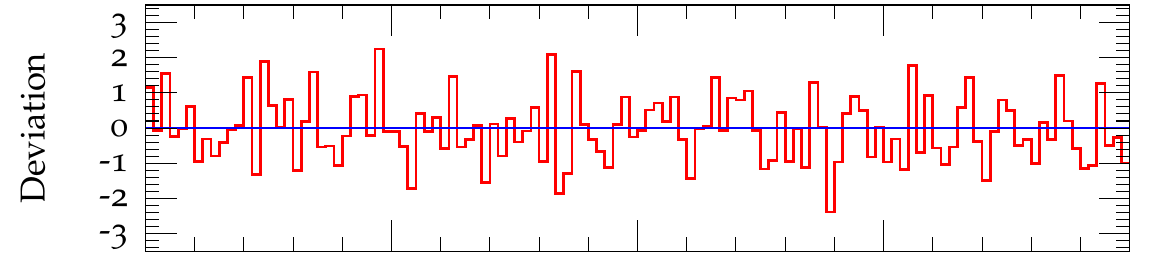}\\[-0.77mm]
    \includegraphics[width=\textwidth]{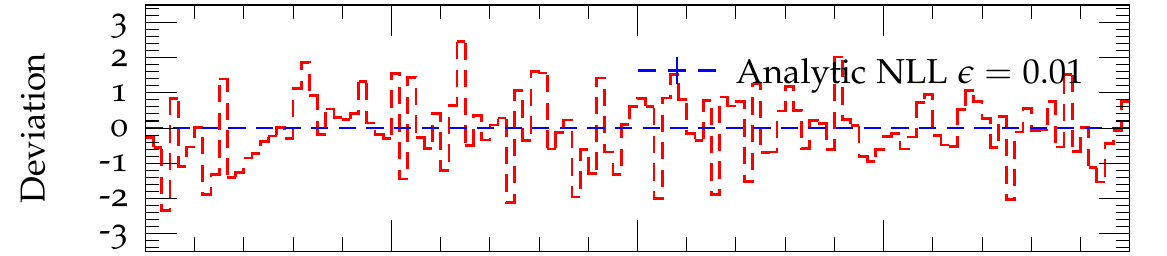}\\[-0.77mm]
    \includegraphics[width=\textwidth]{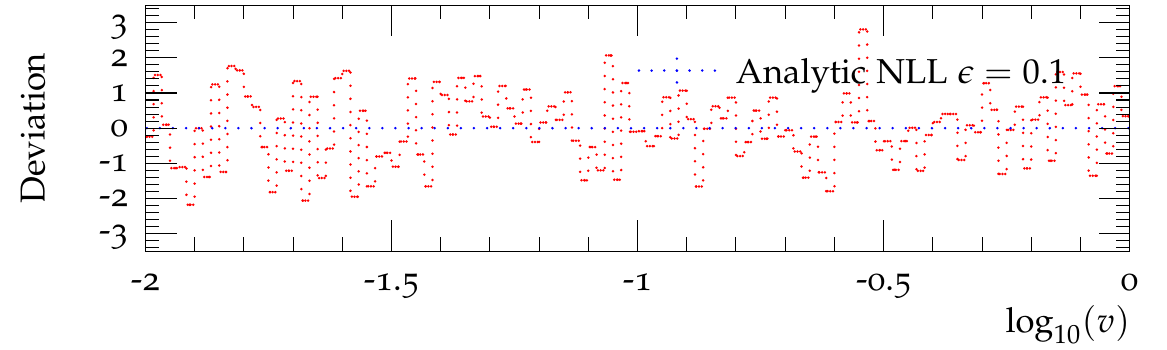}
  \end{minipage}\label{fig:epsilon_tau}}
  \subfigure[~BKS$_{1/2}$]{\begin{minipage}{0.32\textwidth}
    \includegraphics[width=\textwidth]{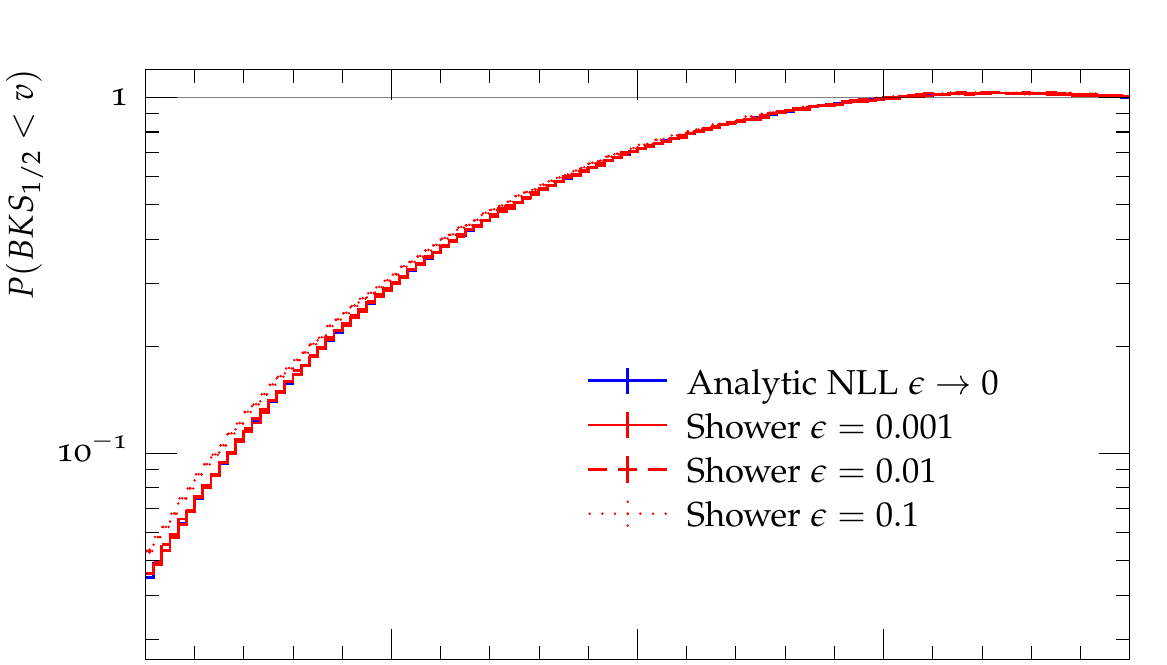}\\[-0.77mm]
    \includegraphics[width=\textwidth]{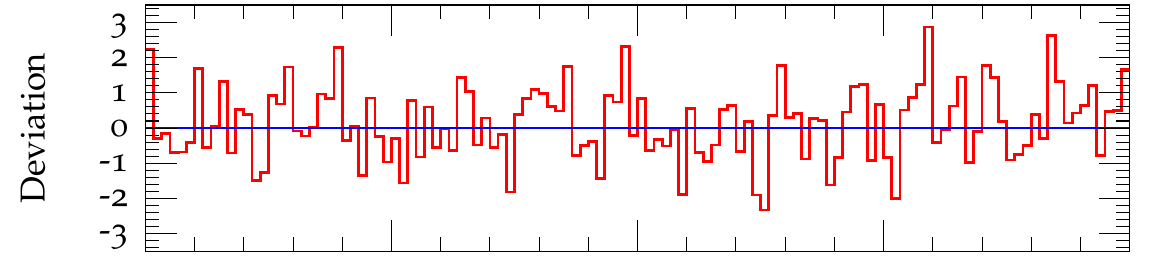}\\[-0.77mm]
    \includegraphics[width=\textwidth]{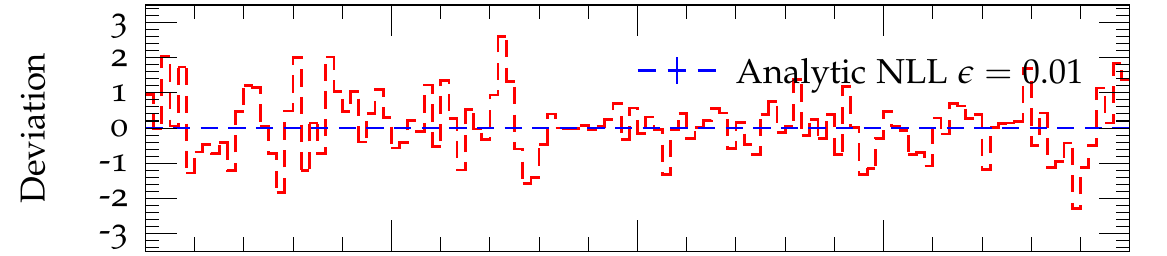}\\[-0.77mm]
    \includegraphics[width=\textwidth]{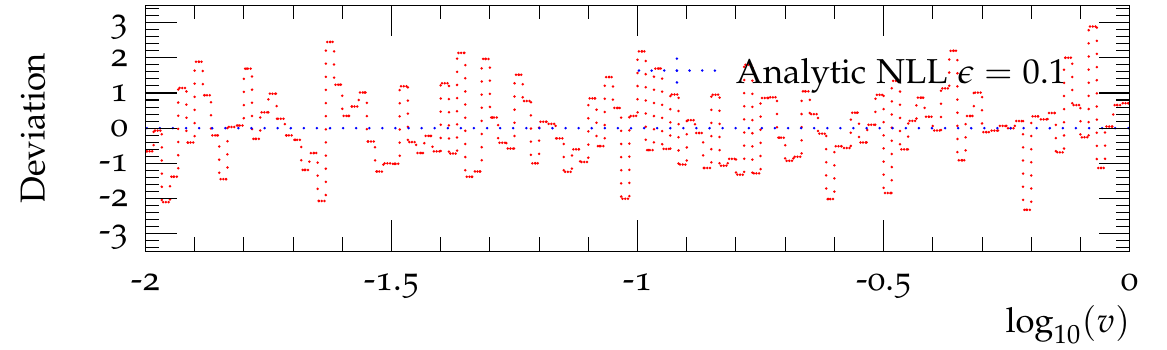}
  \end{minipage}\label{fig:epsilon_bks}}
  \subfigure[~FC$_{1}$]{\begin{minipage}{0.32\textwidth}
    \includegraphics[width=\textwidth]{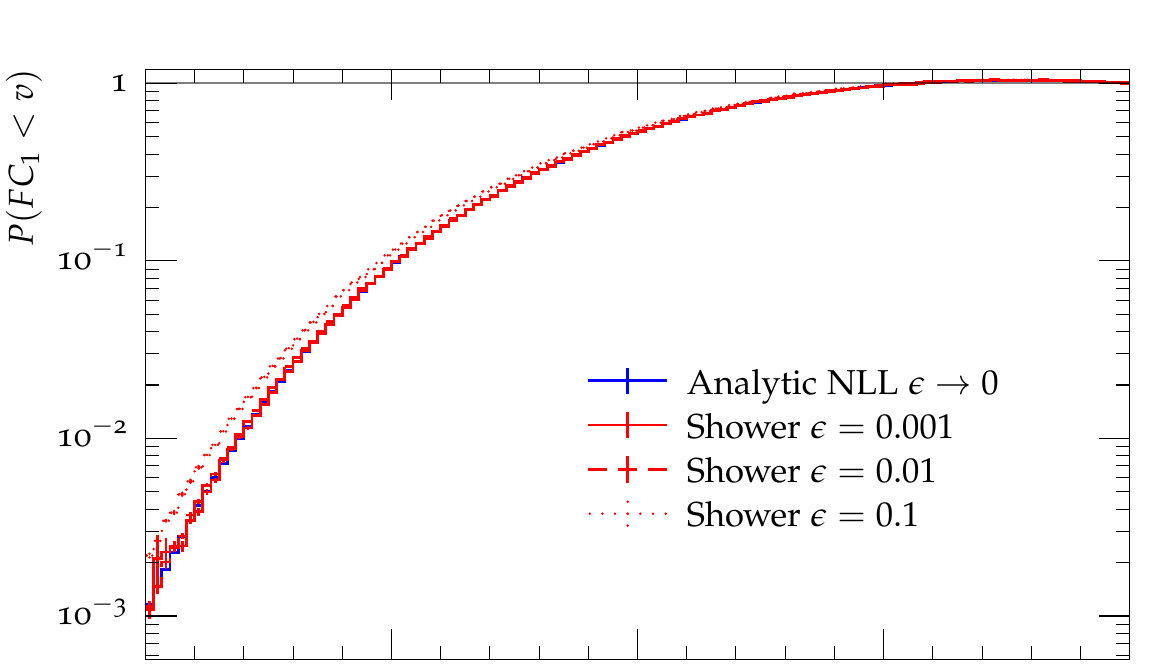}\\[-0.77mm]
    \includegraphics[width=\textwidth]{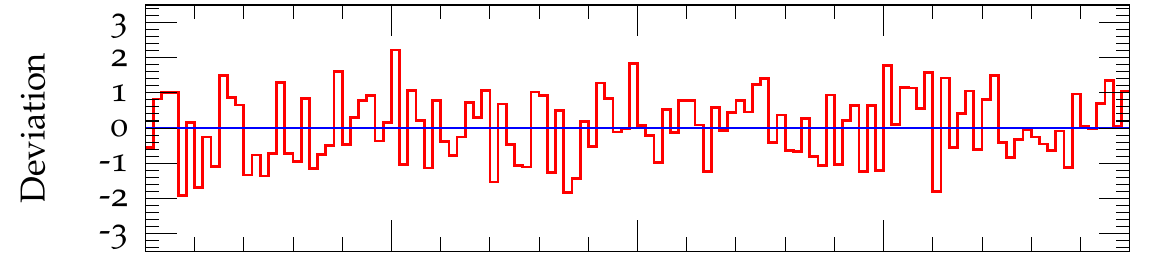}\\[-0.77mm]
    \includegraphics[width=\textwidth]{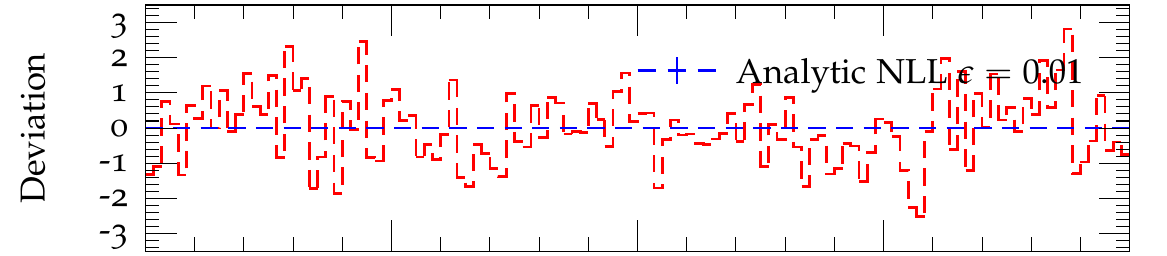}\\[-0.77mm]
    \includegraphics[width=\textwidth]{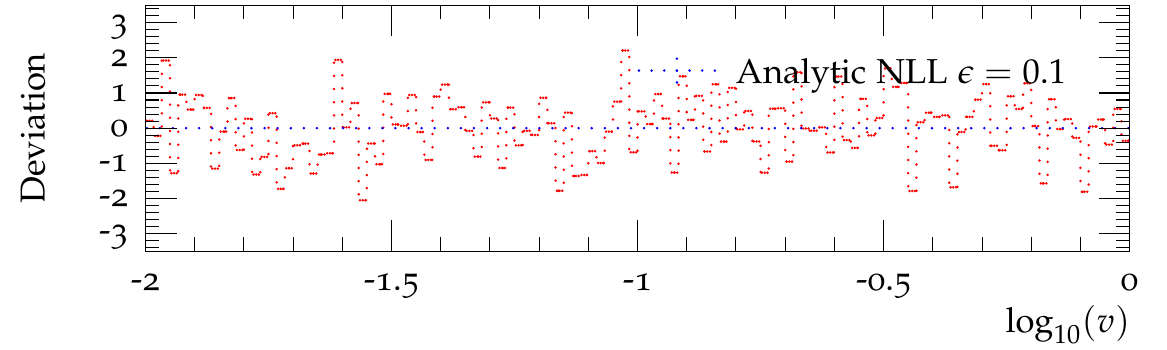}
  \end{minipage}\label{fig:epsilon_fc}}
  \caption{Thrust variable $1-T$ \subref{fig:epsilon_tau}, BKS observable with $x=1/2$ \subref{fig:epsilon_bks}
    and fractional energy correlation with $x=1$ \subref{fig:epsilon_fc} for different values of the cutoff $\epsilon$ in
  Eq.~\eqref{eq:f_function}. The respective analytic results for $\mc{F}_\epsilon(v)$ are used as a reference
  in the ratio plots.}
\label{fig:epsilon}
\end{figure}

For multiple emissions $P_{\rm res}$ explicitly depends on $v$.
We therefore first choose a value for $v$ and then run the parton shower,
implementing the $z$ integration bounds and the scale of the strong coupling as
defined in Tab.~\ref{tab:Parameters}. This is a highly inefficient procedure
to compute the cumulative cross section. If probability was conserved,
the same distribution could be obtained by running the parton shower,
computing $v$, filling the histogram in each bin with lower edge larger than $v$,
and filling the histogram in the bin containing $v$ with weight
$(v_{\rm max}-v)/\Delta v$, where $v_{\rm max}$ is the upper bin edge and
$\Delta v$ is the bin width. This will be the method used to compute the
predictions in Fig.~\ref{fig:FullResult} and Sec.~\ref{sec:DGLAPvsCS}.
While at the level of accuracy we are interested in, it is sufficient
to set the cutoff scale of the parton shower to some numerically small
value in $\xi$, exact agreement with the analytic calculation is expected
only if the calculation is performed for a finite $\epsilon$, and the
parton-shower cutoff is set to $\xi_c = \epsilon v$. We can verify that in
this situation we reproduce the analytic result for finite $\epsilon$ in
Eq.~\eqref{eq:f_function} and investigate the convergence towards the
analytic result for $\epsilon\to 0$. Figure~\ref{fig:epsilon} presents the
corresponding comparison for different values of $\epsilon$ in the case of
the thrust \subref{fig:epsilon_tau}~\cite{Farhi:1977sg}, a BKS observable
\subref{fig:epsilon_bks}~\cite{Berger:2002ig,Berger:2003iw} and
a fractional energy correlation \subref{fig:epsilon_fc}~\cite{Banfi:2004yd}.
The definitions of the observables and related resummation coefficients
are listed in App.~\ref{sec:observables}.

\section{Effects of approximations}
\label{sec:analysis}
This section is dedicated to the detailed investigation of the effects of
local four-momentum conservation and approximations made in the NLL calculation
compared to the parton shower. In order to cover different choices of the parameters
$a$ and $b$, we again present results for the thrust, a BKS observable ($x=1/2$)
and a fractional energy correlation ($x=1$).
All distributions are shown for $Q=91.2~\mathrm{GeV}$, and for a strong coupling
defined by $\alpha_s(Q^2)=0.118$ and a fixed number of flavors, $n_f=5$.
We have cross-checked all of our predictions using two independent Monte-Carlo
implementations based on~\cite{Hoche:2014rga}.

\begin{figure}
  \centering
  \subfigure[~Thrust]{\includegraphics[width=.3\textwidth]{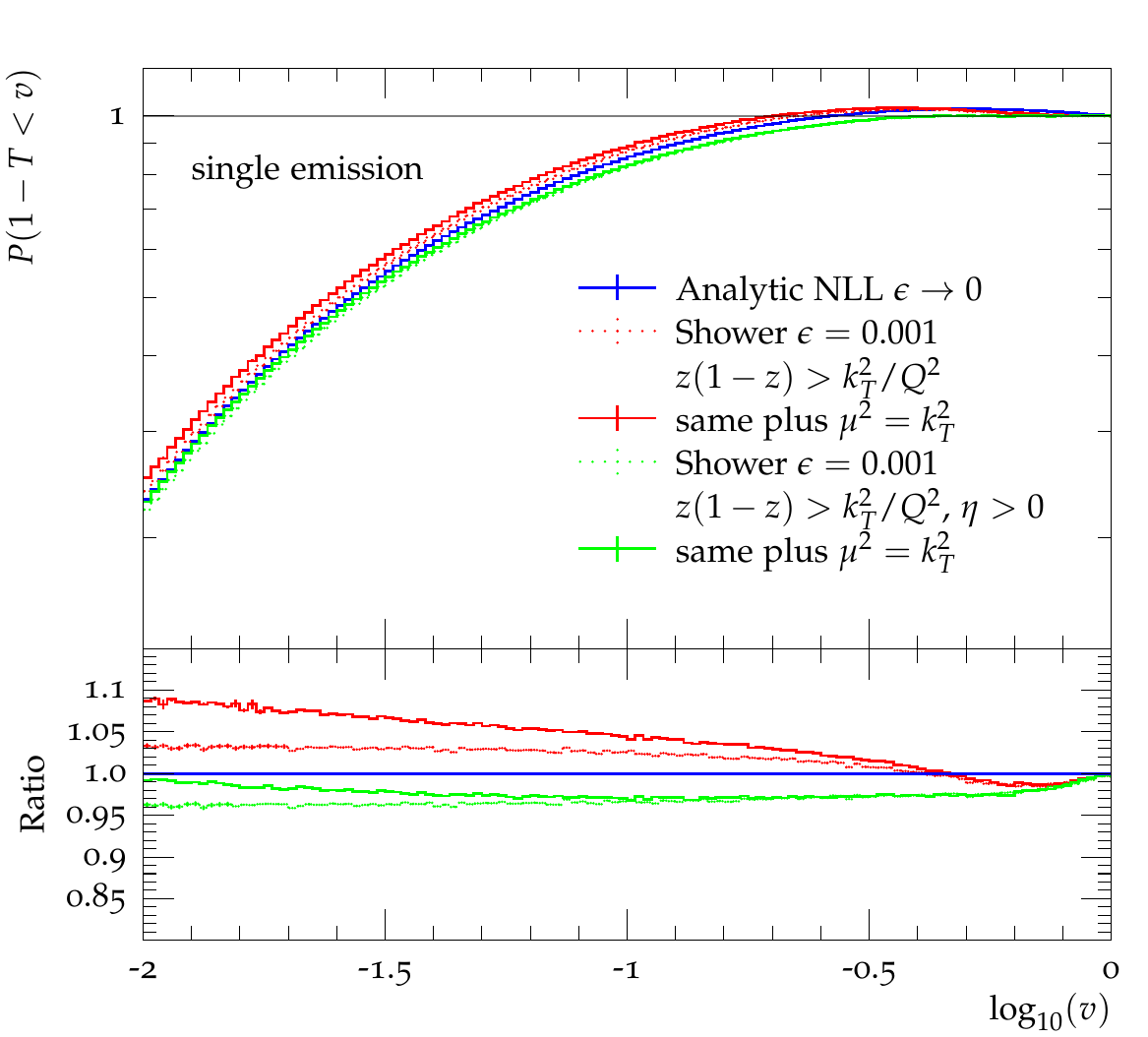}}
  \subfigure[~BKS$_{1/2}$]{\includegraphics[width=.3\textwidth]{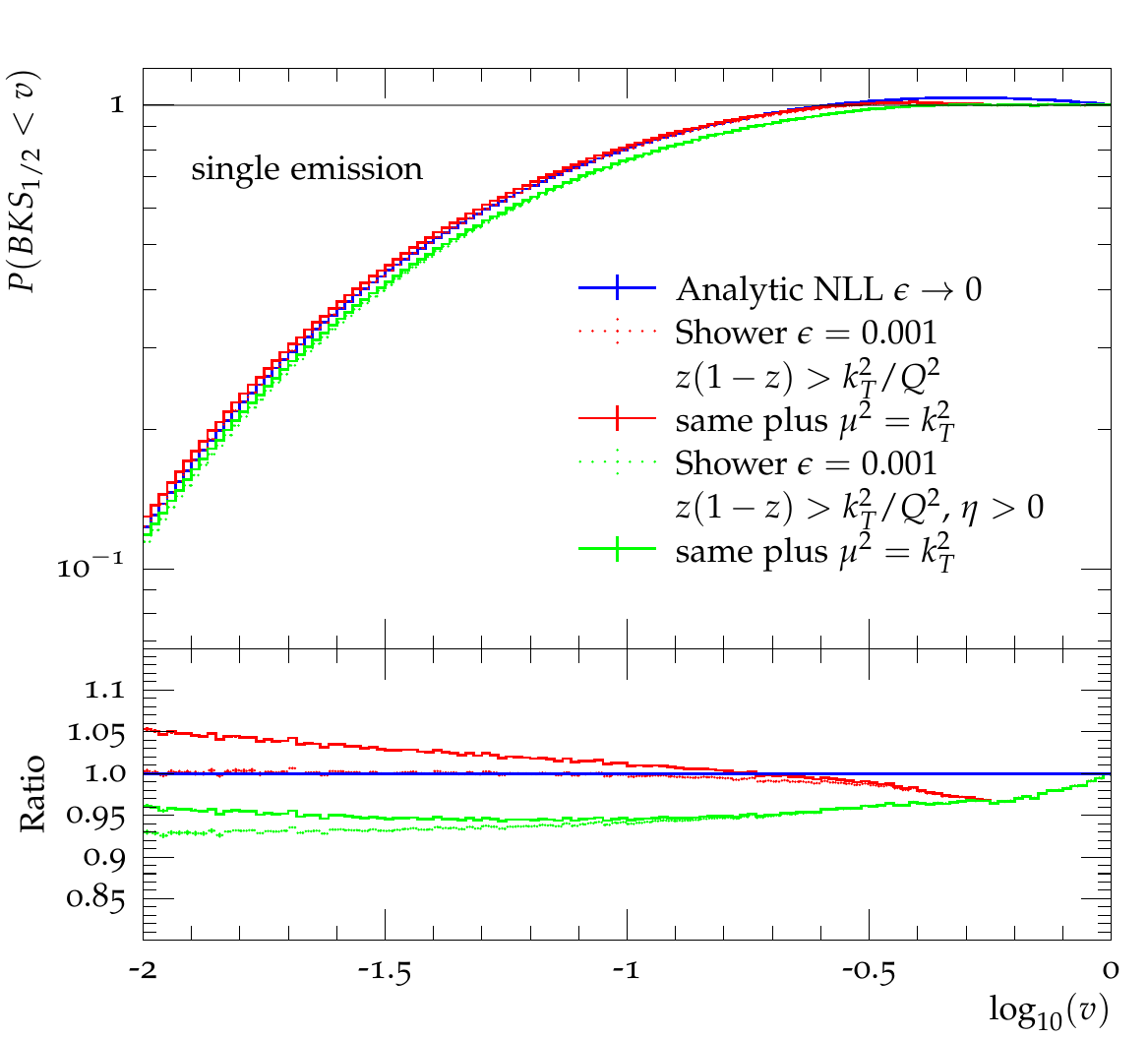}}
  \subfigure[~FC$_{1}$]{\includegraphics[width=.3\textwidth]{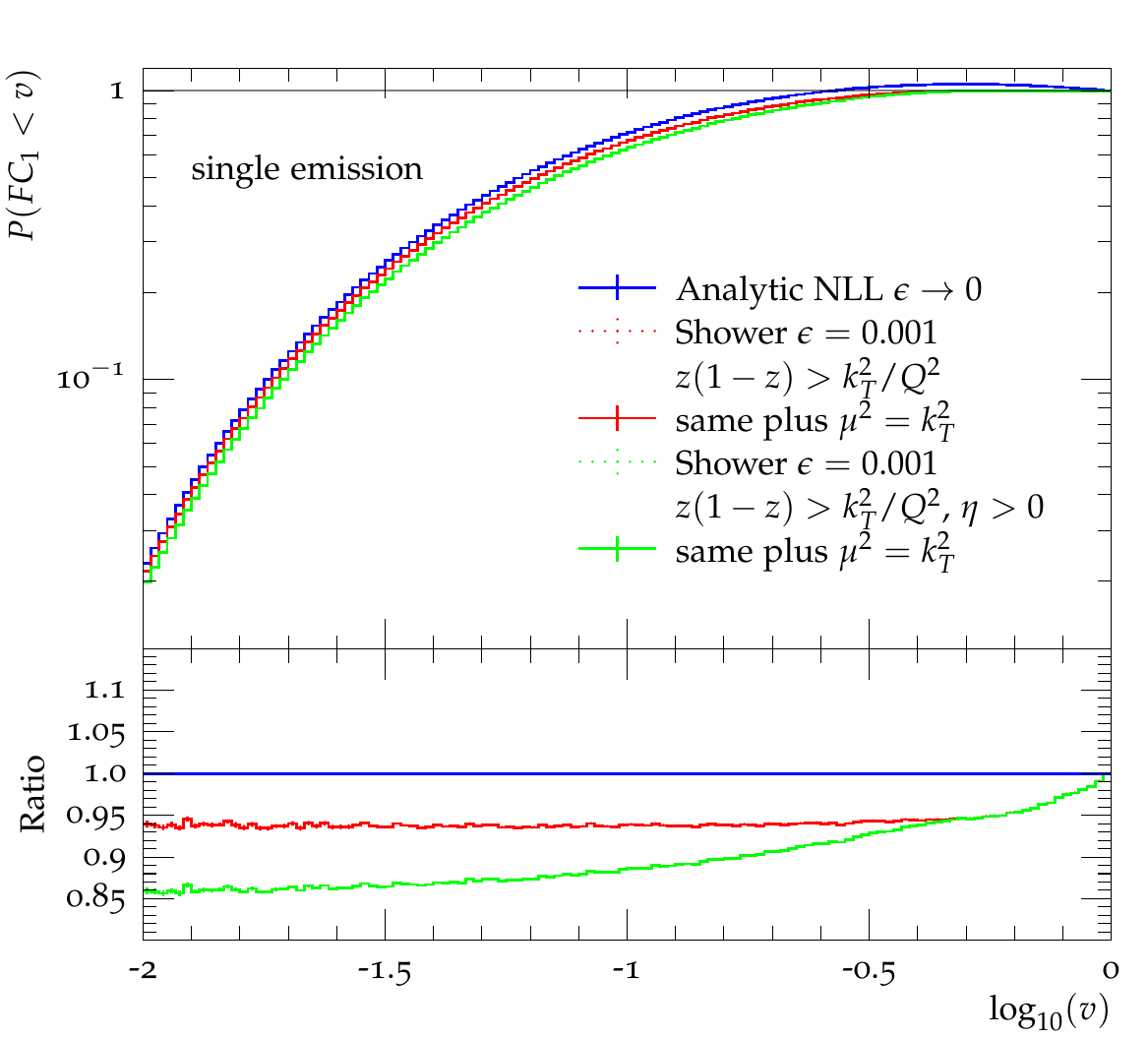}}
  \caption{Effects arising from momentum-conservation in the anti-collinear direction
    and from phase-space sectorization (removal of soft double counting in typical
    parton-shower implementations). Both are effects at the single-emission level,
    impacting terms in $R(v)$, cf.\ Tab.~\ref{tab:Parameters}.}
  \label{fig:kinConstSoftFinite}
\end{figure}

We first investigate constraints arising from momentum conservation in the
anti-collinear direction at single emission level, which reads
\begin{equation}
  Q^2>2p_ip_j=\frac{k_T^2}{z(1-z)}\;.
\end{equation}
This induces both a lower and an upper bound on $z$ given by $z_{\rm min/max}=(1\mp\sqrt{1-4k_T^2/Q^2})/2$.
Figure~\ref{fig:kinConstSoftFinite} shows a comparison between the pure NLL predictions
and those where this constraint has been implemented. The effect on the cumulative
distributions is moderate, about 5\% in the medium and low-$v$ region.
In addition, we investigate the effect of choosing the scale in the collinear term
to be $k_T^2$. This alters the slope of the thrust and BKS$_{1/2}$ distributions
in the small-$v$ region, due to additional sub-leading logarithmic terms in $R(v)$.

The upper bound $z_{\rm max}$ is generally weaker than the constraint arising
from the condition $\eta>0$, listed in Tab.~\ref{tab:Parameters}.
Figure~\ref{fig:kinConstSoftFinite} displays the additional effect on the NLL prediction
when this constraint is applied in form of $z_{\lessgtr v,\rm coll}^{\rm max}$
as used in typical parton showers (cf.\ Eq.~\eqref{eq:single_emission_ps}).
The effects are about 10\% on all observables, and they lower the prediction
for $\Sigma(v)$ due to an increased branching probability.
Again, we also investigate the effect of choosing the scale in the collinear term
to be $k_T^2$, which generates the same slope differences at small $v$ observed before.

\begin{figure}
  \centering
  \subfigure[~Thrust]{\includegraphics[width=.3\textwidth]{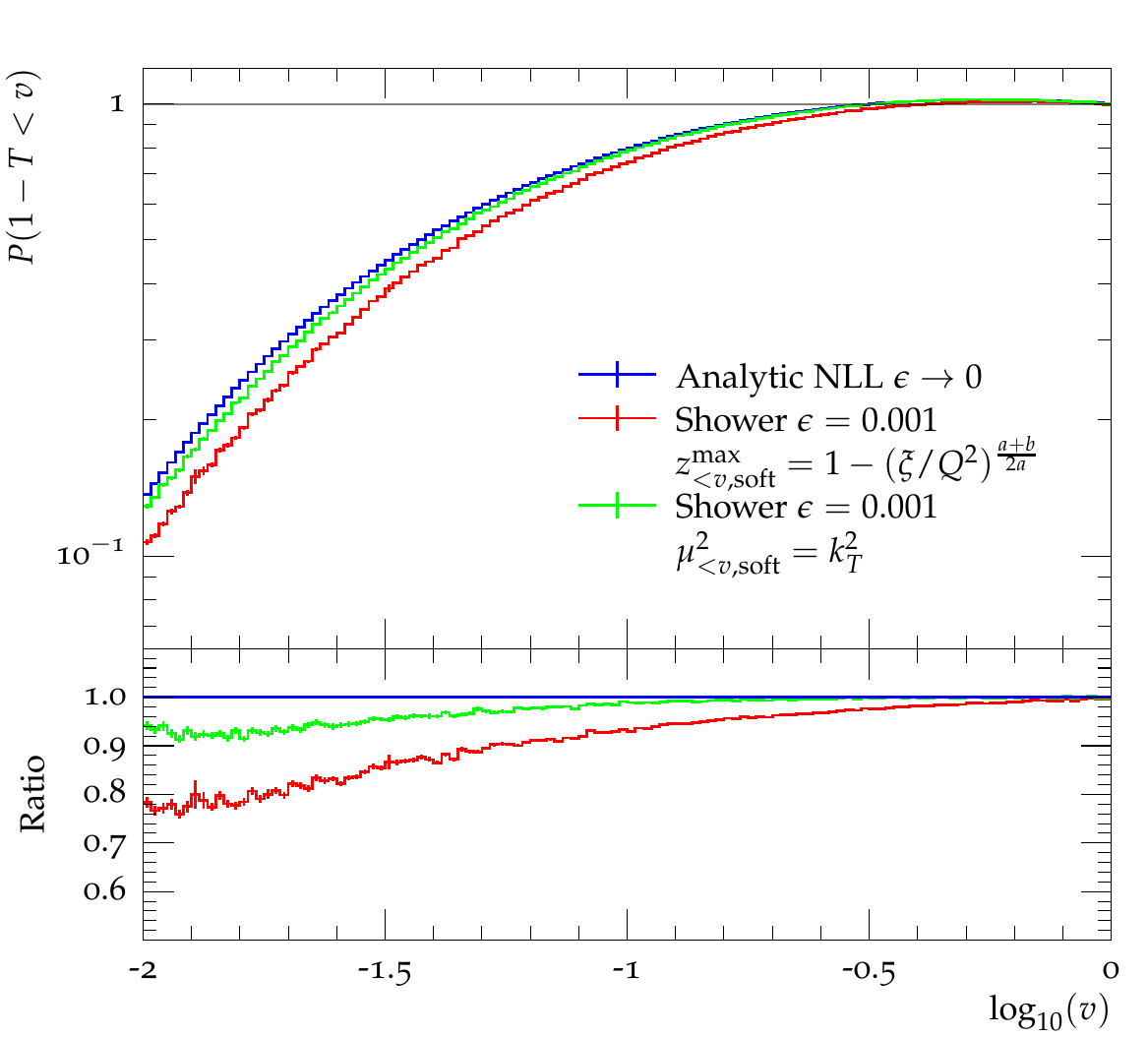}}
  \subfigure[~BKS$_{1/2}$]{\includegraphics[width=.3\textwidth]{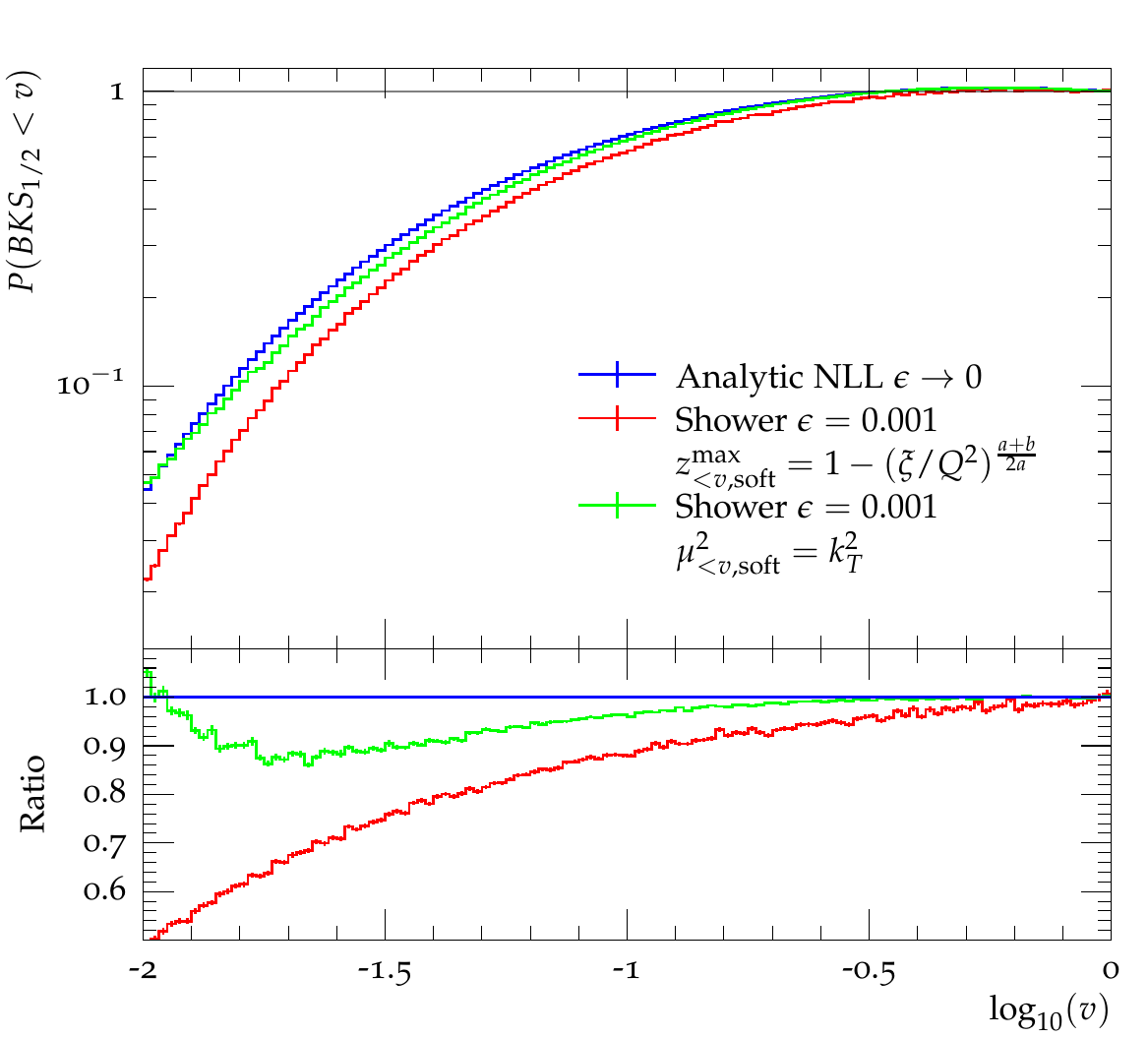}}
  \subfigure[~FC$_{1}$]{\includegraphics[width=.3\textwidth]{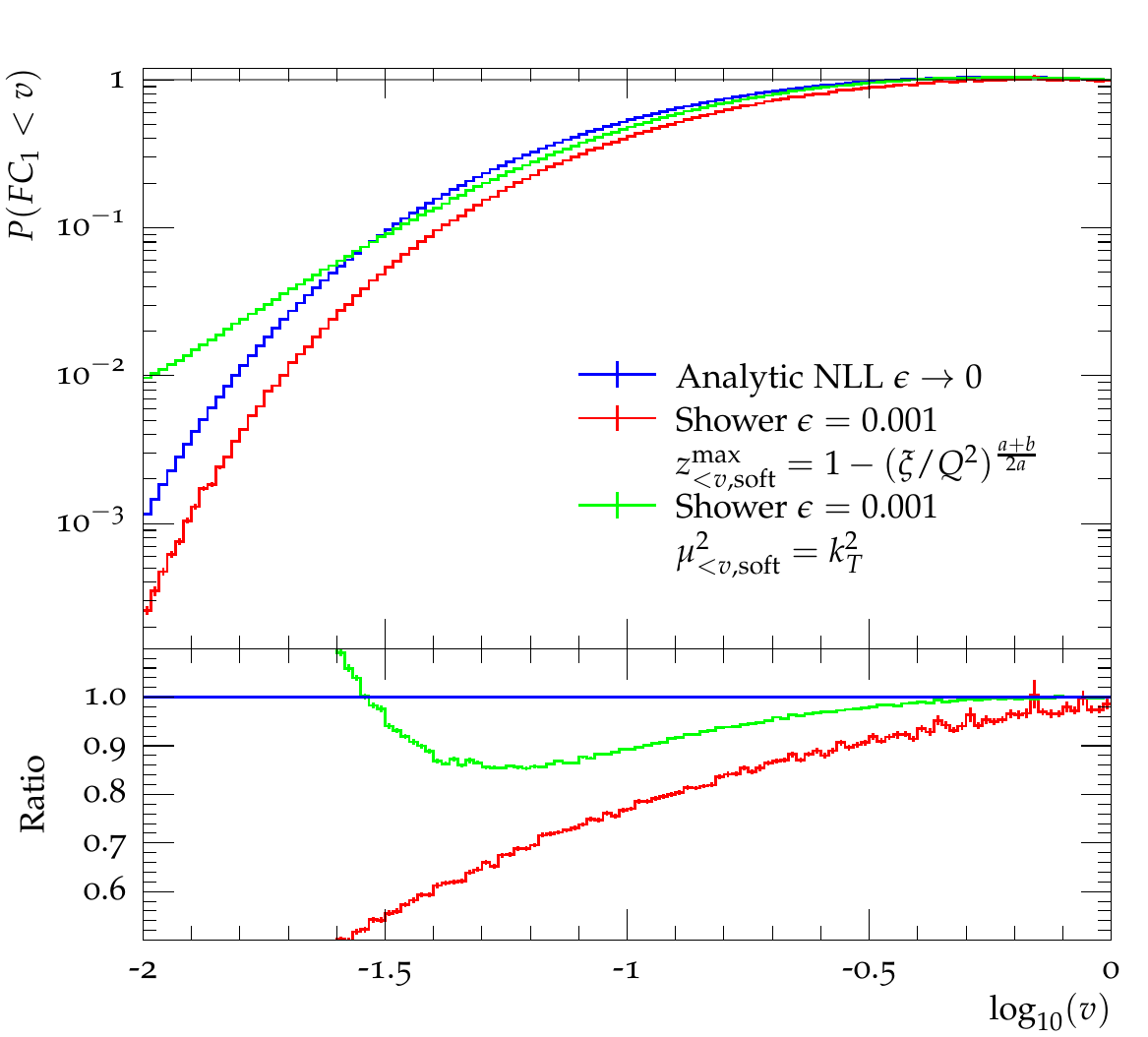}}
  \caption{Effect of replacing the constraint $z<1-v^{1/a}$ for $\xi<Q^2v^{2/(a+b)}$
    by the phase-space sectorization constraint, $\eta>0$ and effects arising from 
    the evaluation of the strong coupling at $k_T^2$ with $k_T^{\rm min}=0.5~{\rm GeV}$.}
  \label{fig:kinConstSoftEnhanced}
\end{figure}

Next we investigate the effect of lifting the restriction on the $z$
integration in the calculation of $\mathcal{F}(v)$, i.e.\ removing the constraint $z<1-v^{1/a}$
if $\xi<Q^2v^{2/(a+b)}$ and replacing it by the constraint $\eta>0$.
In this case $R'(v)$ must be computed down to very small scales in Eq.~\eqref{eq:f_function},
(except for FC$_1$) and it becomes mandatory to introduce an additional cutoff, as one would
otherwise need to evaluate $\alpha_s$ at values where perturbation theory is no longer valid.
We choose to implement this by adding the requirement $k_T^{\rm min}=0.5~{\rm GeV}$.
The difference to the pure NLL result is shown in Fig.~\ref{fig:kinConstSoftEnhanced}.
Independent of the observable, this change is one of the largest differences observed
in this study. The large relative difference between the pure NLL result and
the modified prediction at small $v$ shows that sub-leading logarithmic effects
become important.

We also study the effect originating in the evaluation of the running coupling
at $Q^2v^{2/(a+b)}(1-z)^{2b/(a+b)}$ if $\xi<Q^2v^{2/(a+b)}$. Again we implement
the constraint $k_T^{\rm min}=0.5~{\rm GeV}$. Figure~\ref{fig:kinConstSoftEnhanced}
shows that the predictions for all observables exhibit relatively large changes.
They also show convergence issues at small values, which arise from the lower cutoff
in $k_T$, leading to an insufficient sampling of $\mc{F}$ at higher number of
emissions. This effect is most pronounced in FC$_1$, where it starts to appear
around $10^{-1.5}$. Note, however, that practical measurements of FC$_1$ would
be impacted by non-perturbative corrections in this regime. The problem is
therefore purely academic in nature, hence we do not attempt to solve it here.

\begin{figure}[t]
  \centering
  \subfigure[~Thrust]{\includegraphics[width=.3\textwidth]{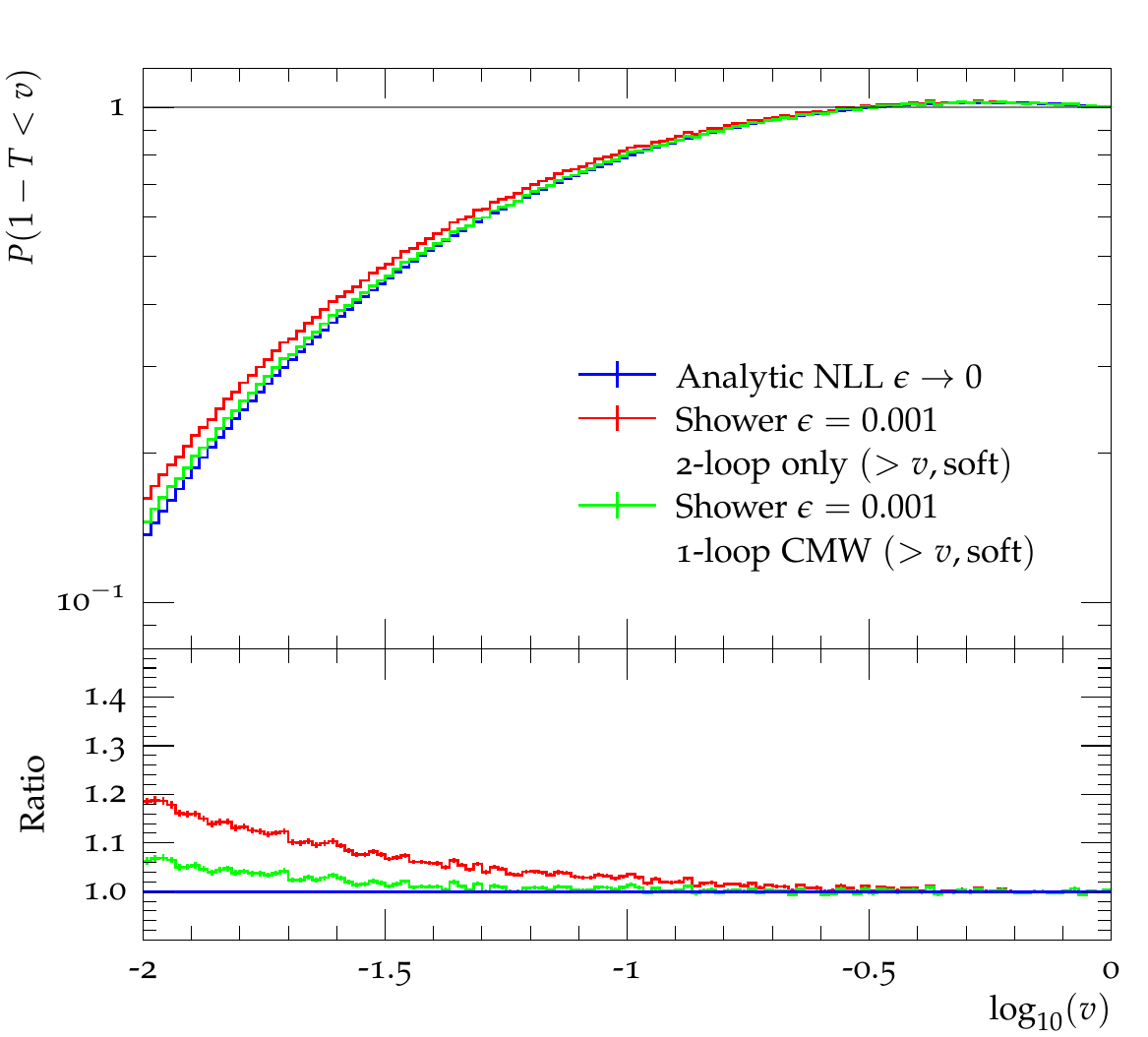}}
  \subfigure[~BKS$_{1/2}$]{\includegraphics[width=.3\textwidth]{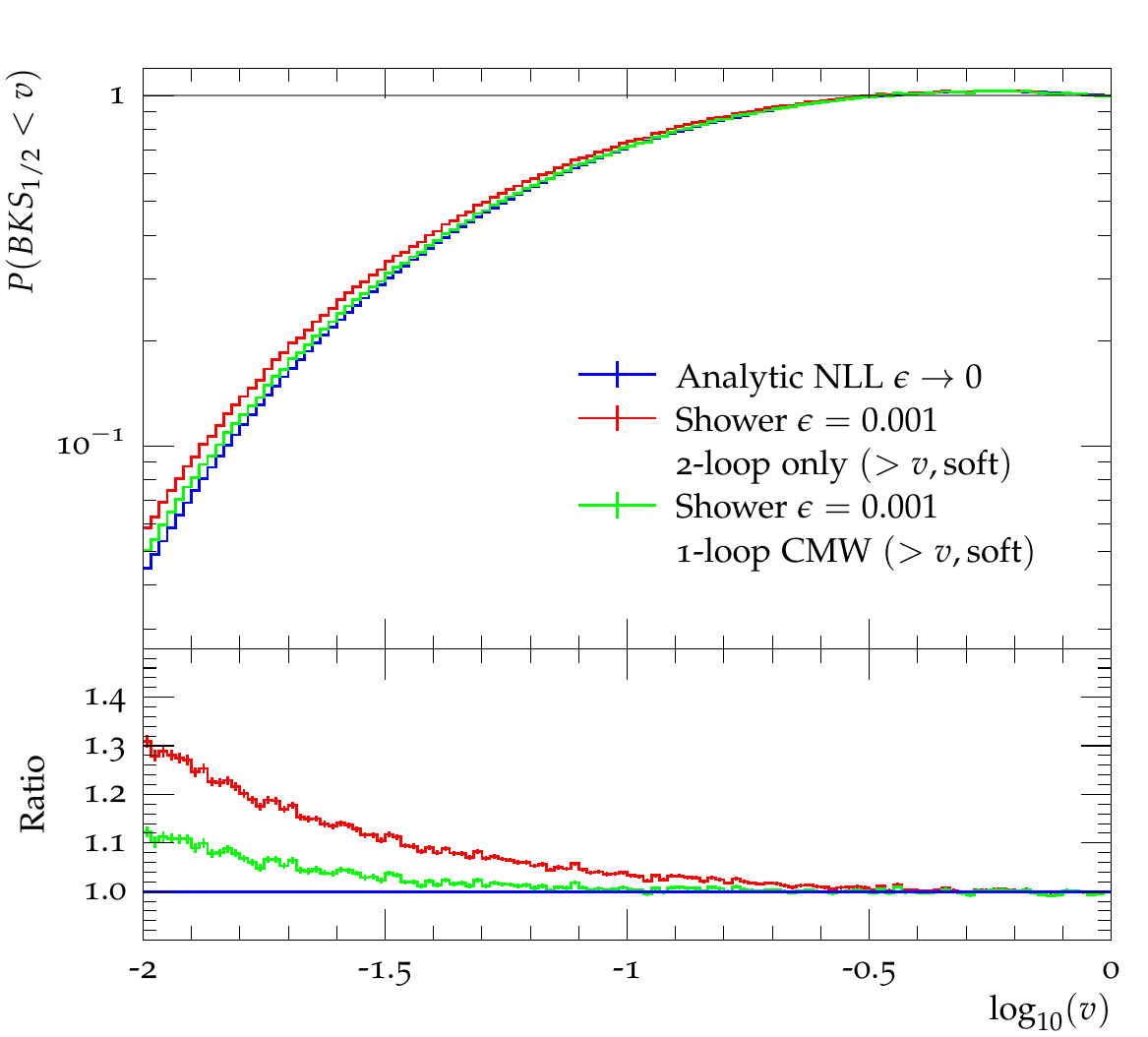}}
  \subfigure[~FC$_{1}$]{\includegraphics[width=.3\textwidth]{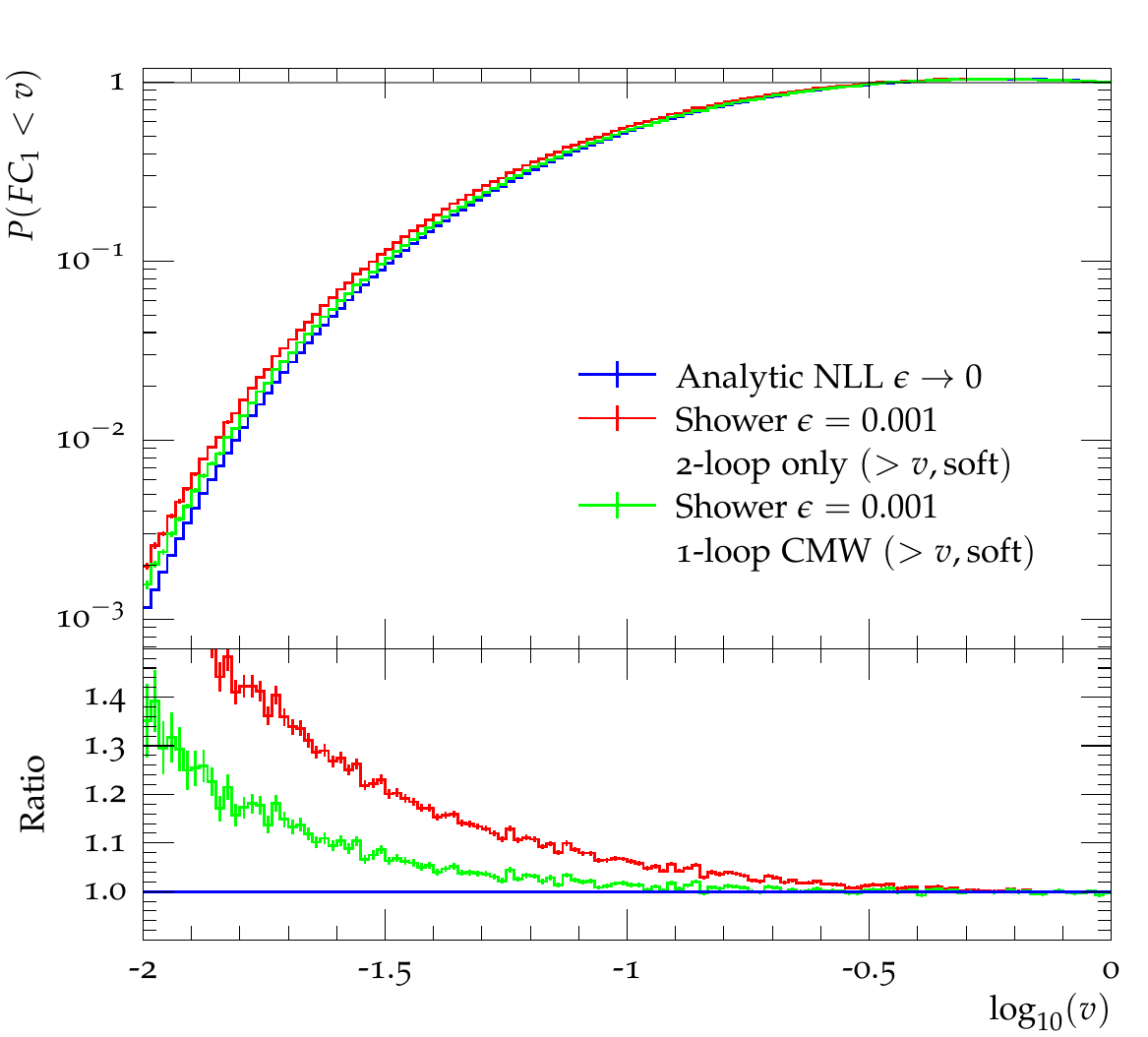}}
  \caption{Effects of replacing 2-loop CMW running of $\alpha_s$ in the leading terms
    of the NLL result. The red line is computed without the CMW scheme,
    while the green line is computed by using the CMW scheme at 1-loop.}\label{fig:CMW} 
\end{figure}

\begin{figure}
  \centering
  \subfigure[~Thrust]{\includegraphics[width=.3\textwidth]{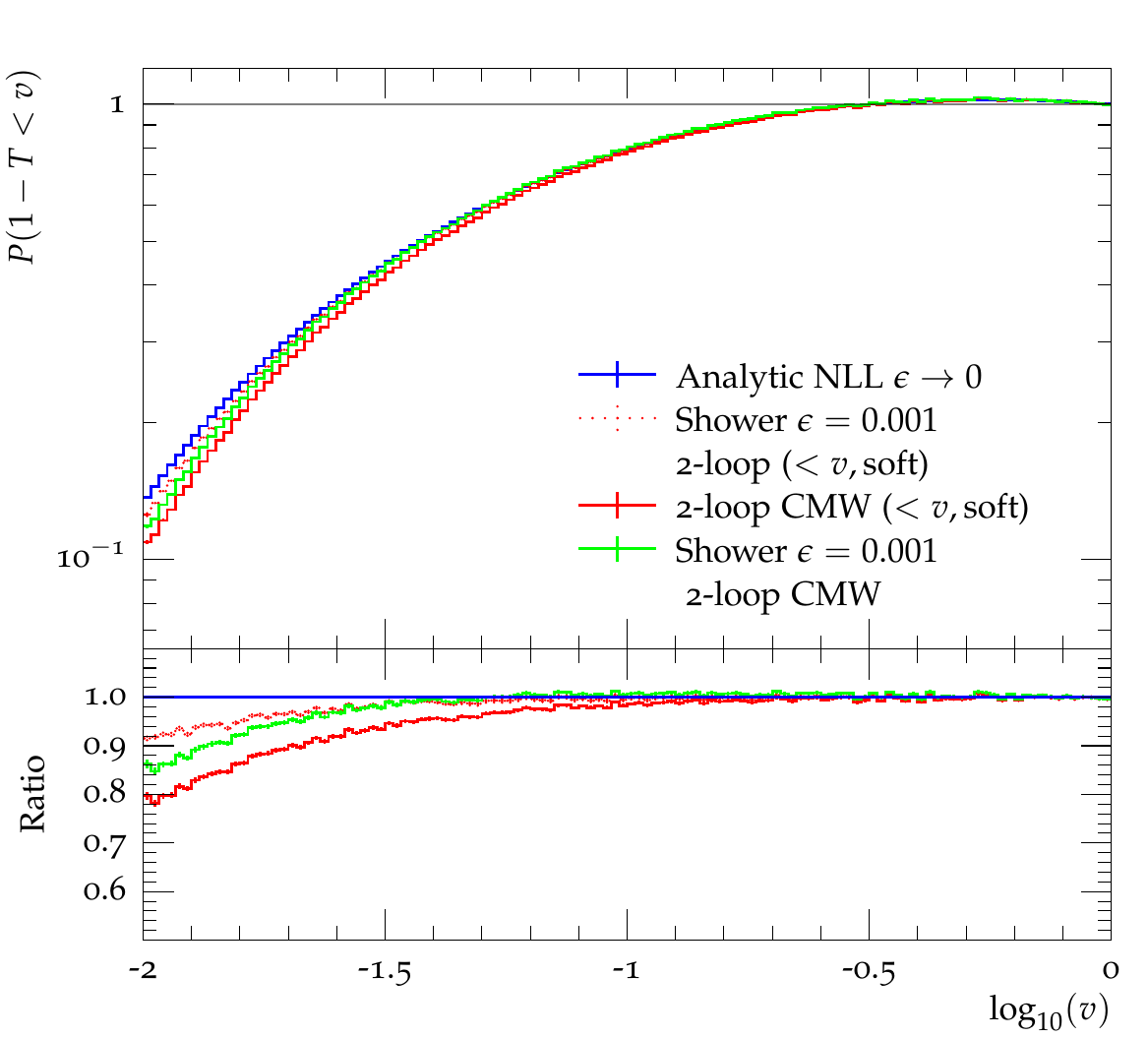}}
  \subfigure[~BKS$_{1/2}$]{\includegraphics[width=.3\textwidth]{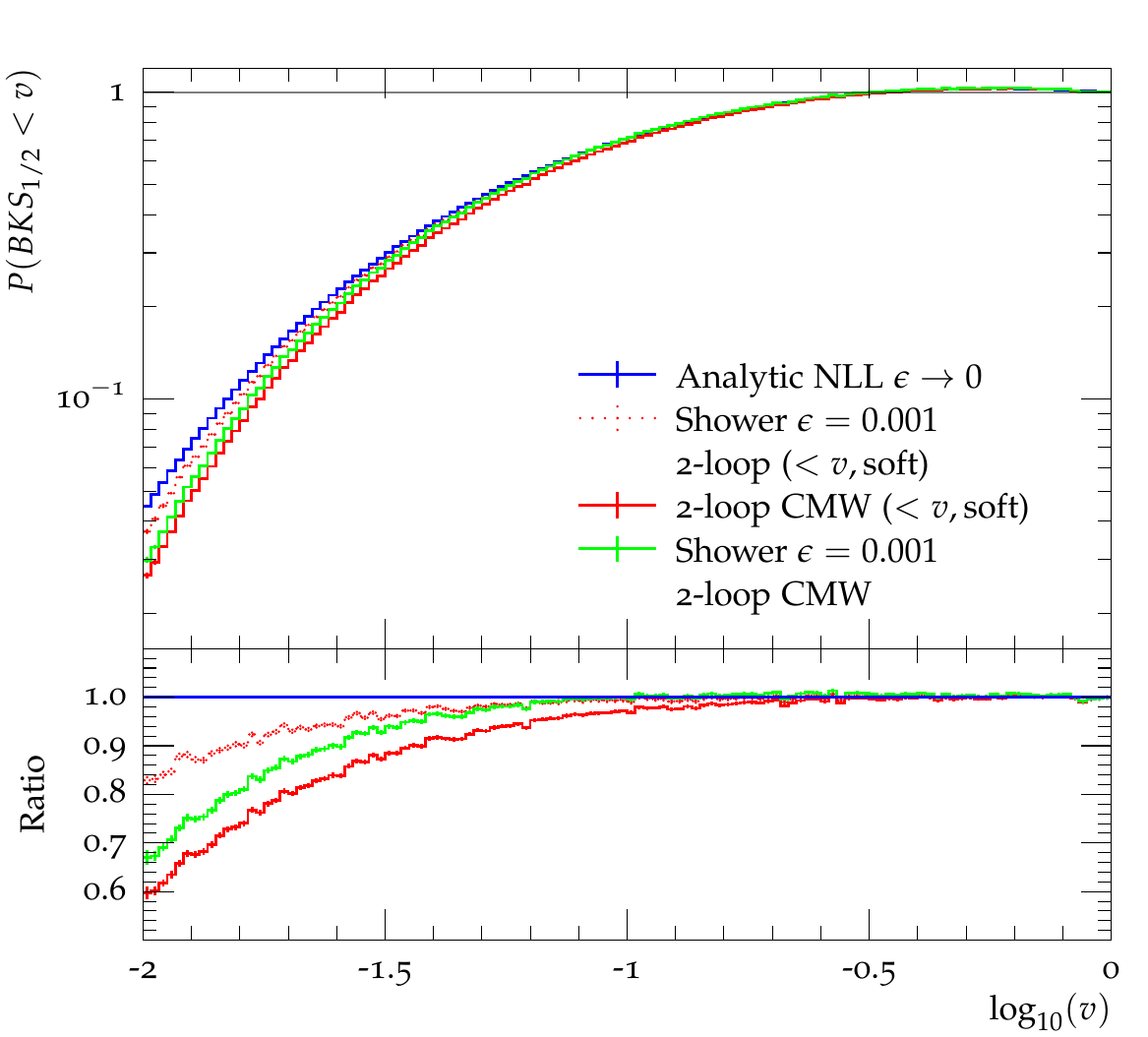}}
  \subfigure[~FC$_{1}$]{\includegraphics[width=.3\textwidth]{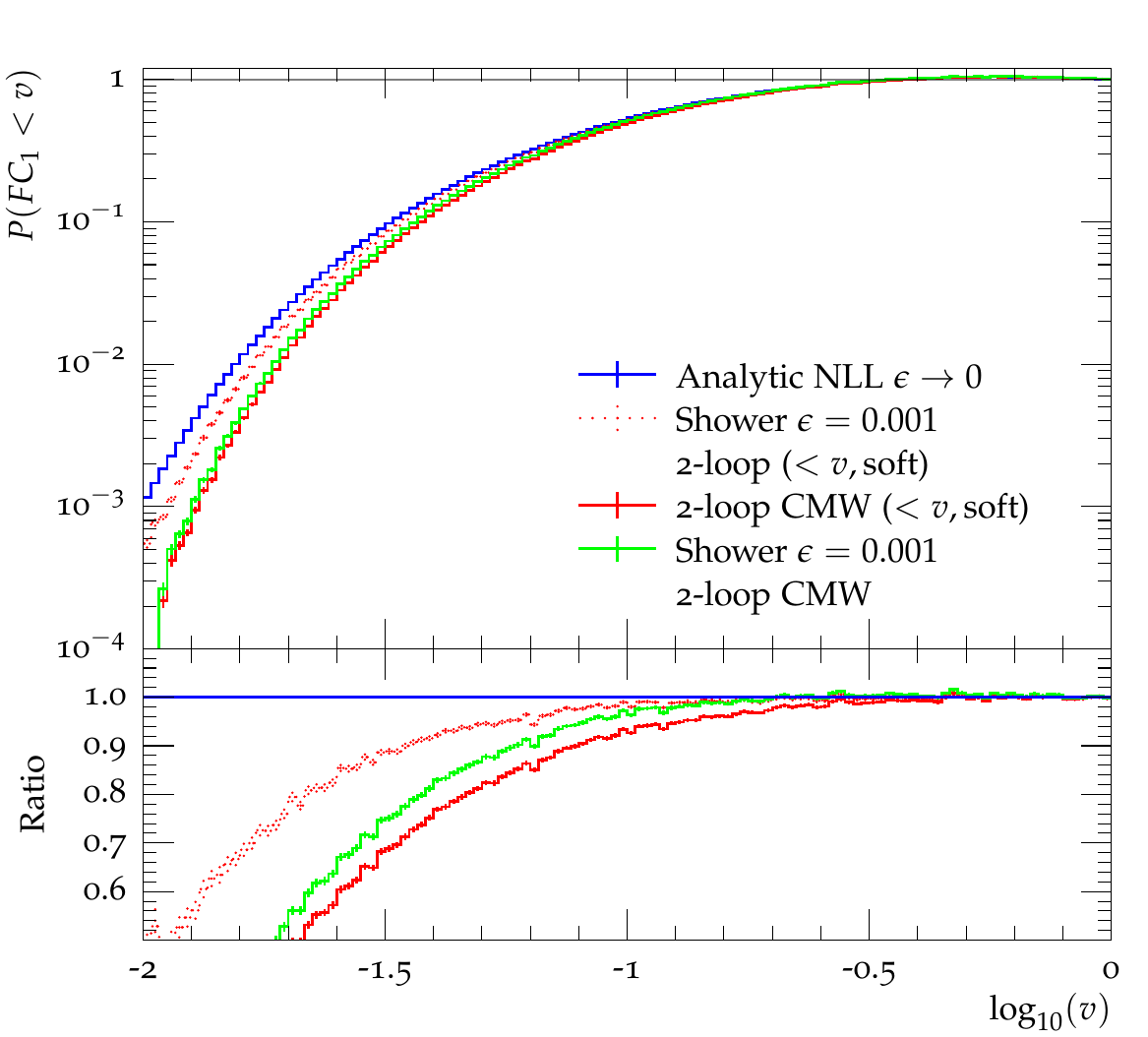}}
  \caption{Effects of replacing 1-loop by 2-loop CMW running of $\alpha_s$
    in the sub-leading terms of the NLL result. The red line is computed 
    by making the replacements only in the soft-enhanced term if $\xi<Q^2v^{2/(a+b)}$,
    the red dotted line corresponds to not using the CMW scheme in this region.}\label{fig:AlphaS} 
\end{figure}

Formally the CMW scheme is the key to achieving NLL accuracy in a parton-shower
computation of the observables considered here~\cite{Catani:1990rr}. The numerical
impact on $R(v)$ is investigated in Fig.~\ref{fig:CMW}.
Figure~\ref{fig:AlphaS} displays the effect of replacing 1-loop by 2-loop running
couplings and of using the CMW scheme in sub-leading terms of the NLL calculation 
(cf.\ Tab.~\ref{tab:Parameters}). The red line is computed by making the replacements
only in the soft-enhanced part of the splitting function for $\xi<Q^2v^{2/(a+b)}$,
and the red dotted line corresponds to not using the CMW scheme if $\xi<Q^2v^{2/(a+b)}$.
It is evident that the effects are sizable over most of the observable range, 
and most pronounced at small $v$. The use of the CMW scheme has the biggest impact.
Note in particular that not using the CMW scheme in the computation of $\mc{F}(v)$
has nearly the same impact as not using the CMW scheme in the computation of $R(v)$.

\begin{figure}
  \centering
  \subfigure[~Thrust]{\includegraphics[width=.3\textwidth]{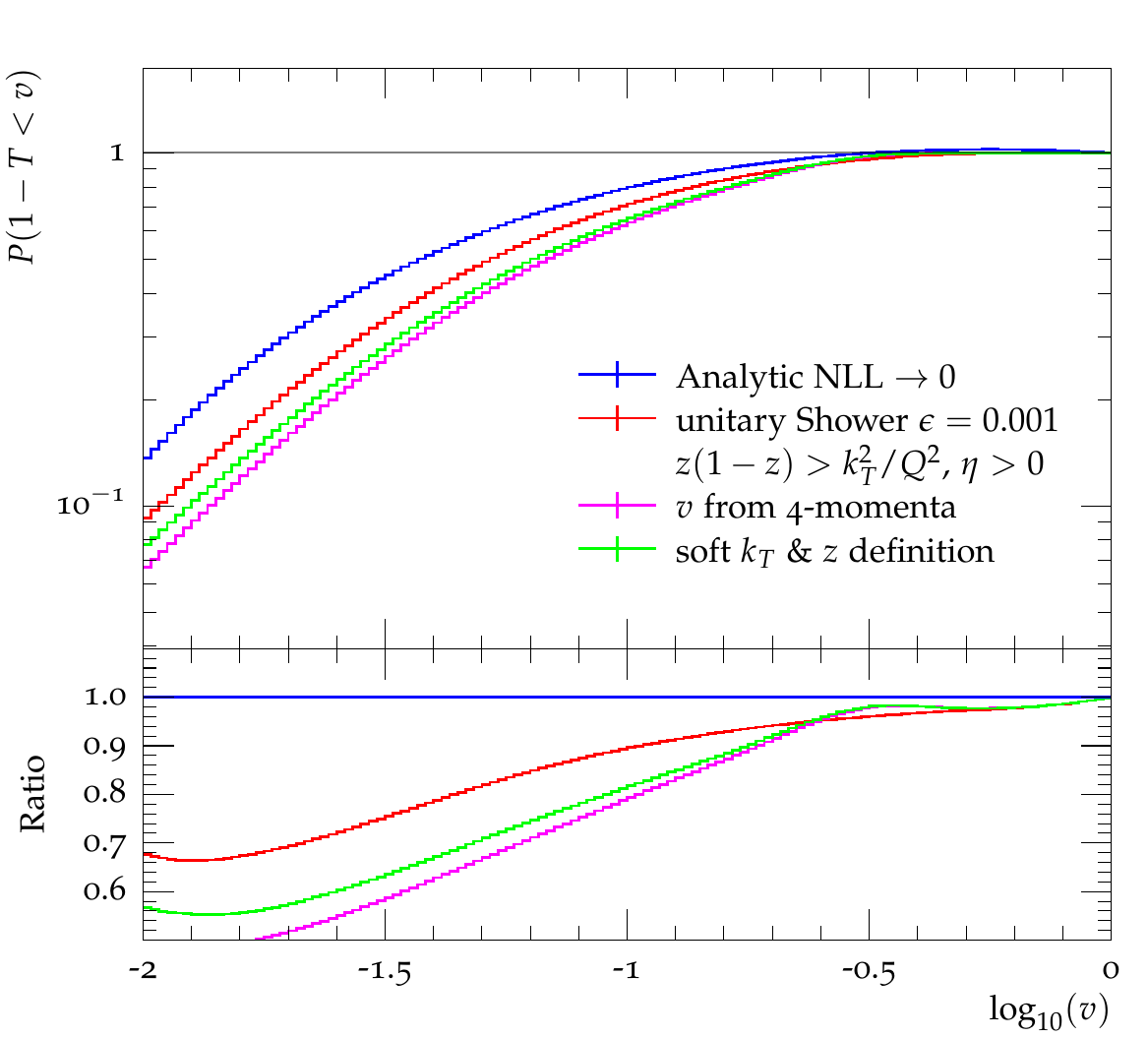}}
  \subfigure[~BKS$_{1/2}$]{\includegraphics[width=.3\textwidth]{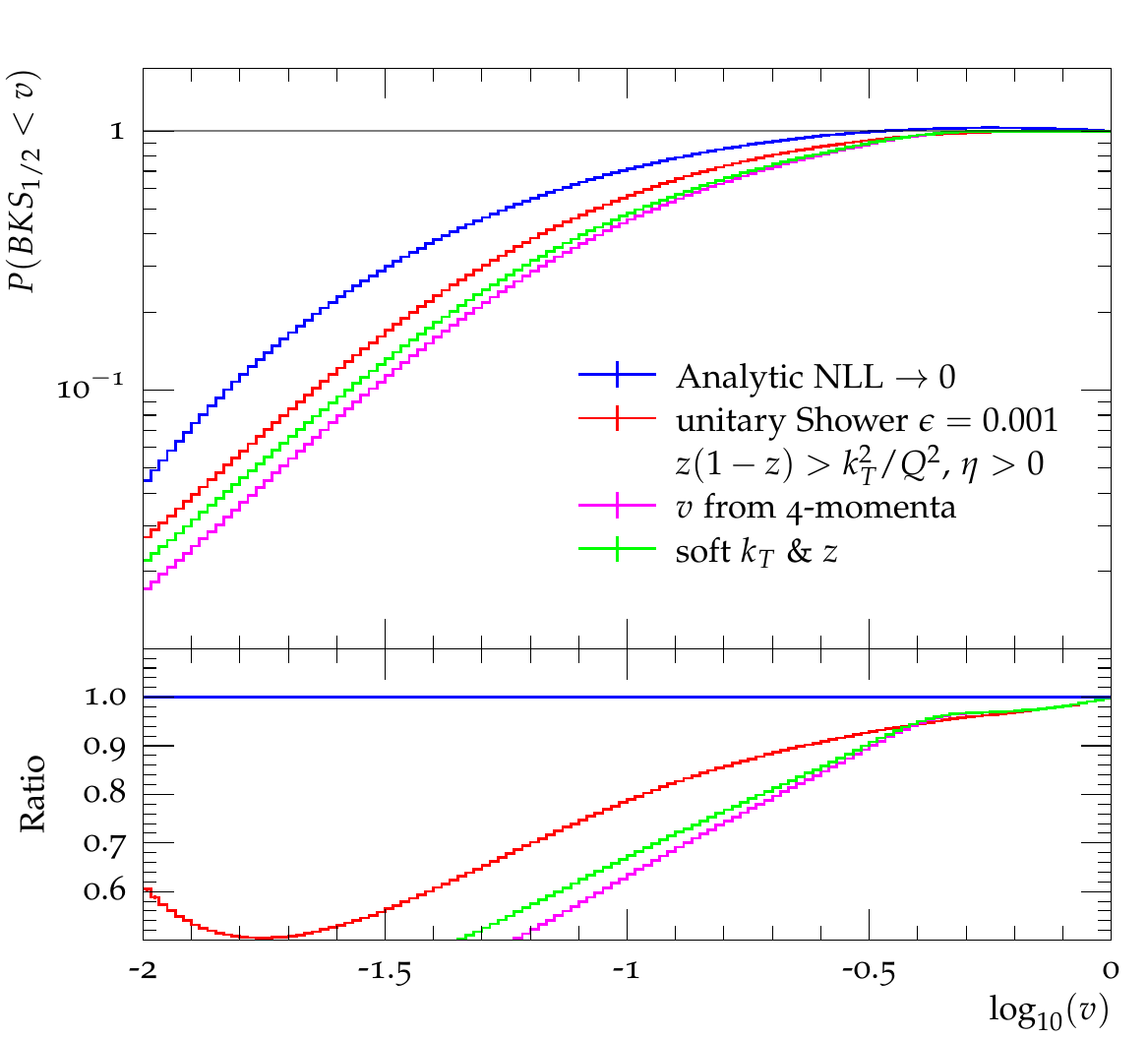}}
  \subfigure[~FC$_{1}$]{\includegraphics[width=.3\textwidth]{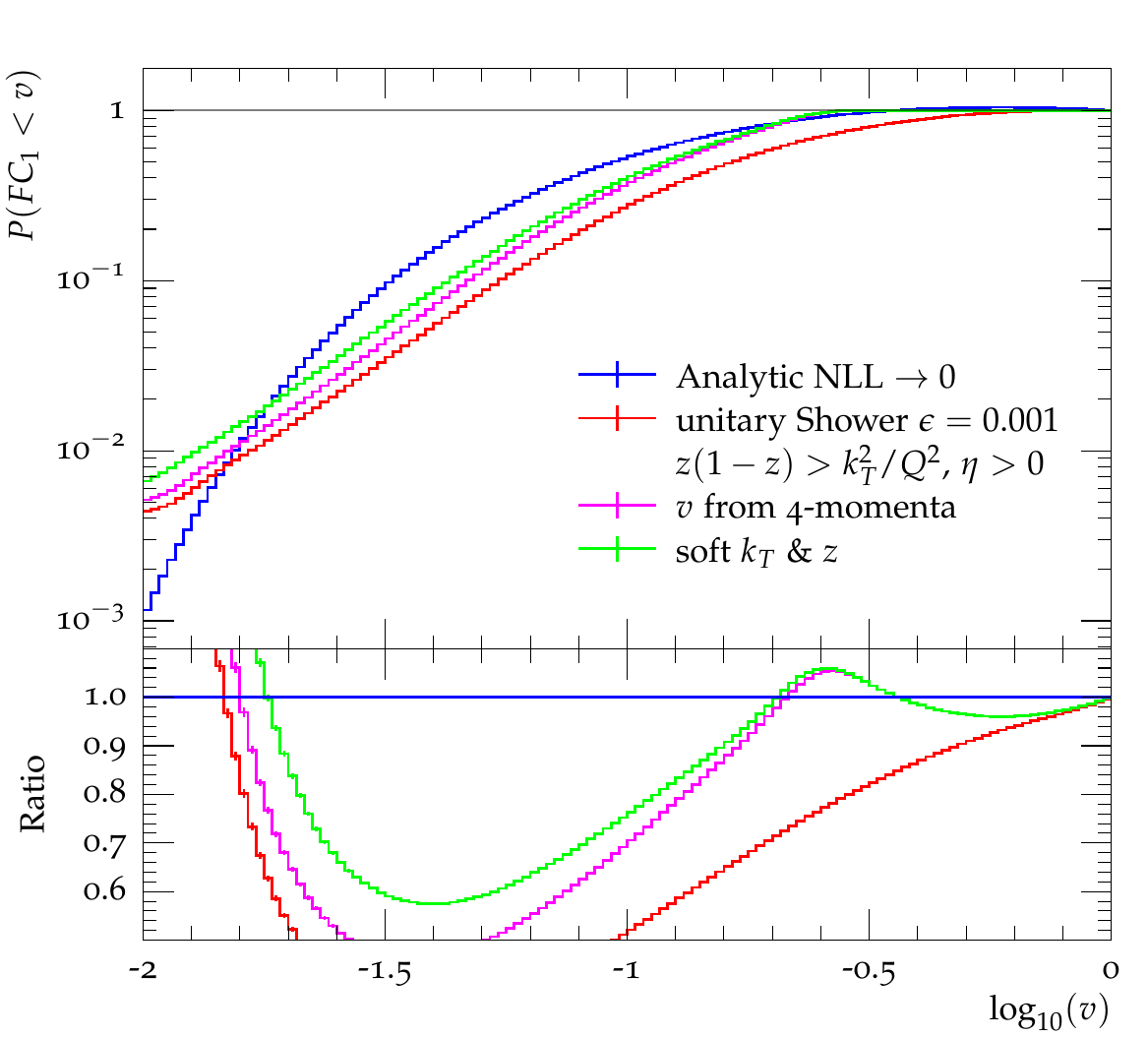}}
  \caption{Comparison of pure NLL resummation and plain DGLAP parton shower, effects of
    approximating the observable compared to exact calculation using four-momenta
    and evolution in dipole-$k_T$.}
  \label{fig:FullResult} 
\end{figure}

Figure~\ref{fig:FullResult} shows the cumulative effect of all changes discussed so far.
In addition we present results from a simulation where the observable is computed using its
definition in terms of four-momenta rather than using the soft approximation in
Eq.~\eqref{eq:V_approx} (see App.~\ref{sec:observables} for details).
In this context it becomes important to take into account
that emissions away from the strict soft limit inevitably change the momenta of the
hard partons. Subsequent emissions are then computed based on the momenta of the
quark lines with recoil effects taken into account. This can have a significant impact on the result,
depending on the precise definition of the transverse momentum and momentum fraction.
The magenta line in Fig.~\ref{fig:FullResult} corresponds to the conventions
of~\cite{Catani:1996vz}, while the green line corresponds to the conventions of~\cite{Hoche:2015sya}.
In the latter case the transverse momentum coincides with Eq.~(2.5) of~\cite{Banfi:2004yd}.\footnote{
  Note that the constraint $z(1-z)>k_T^2/Q^2$ arising from minus momentum conservation
  applies to this definition in the case of final-state emitter with final-state spectator,
  such that the results in Fig.~\ref{fig:kinConstSoftFinite} remain valid.}
Note that the phase-space sectorization constraint, $\eta>0$, generates a different 
restriction on $z$ once recoil is taken into account, and that this condition depends
on the choice of evolution and splitting variable.

\section{Comparison with a dipole-like parton shower}
\label{sec:DGLAPvsCS}
This section presents a comparison of our previous results with predictions from
a dipole-like parton shower. In such parton showers the soft enhanced part of the
collinear splitting function is typically replaced by a partial fraction of the
soft eikonal matched to the collinear limit~\cite{Ellis:1980wv,Catani:1996vz}.
At the same time, the phase-space sectorization is removed, i.e.\ the restriction
$\eta>0$ is lifted. A complete description of the parton-shower algorithm employed
here can be found in~\cite{Hoche:2015sya}.

\begin{figure}
  \centering
  \subfigure[~Thrust]{\includegraphics[width=.3\textwidth]{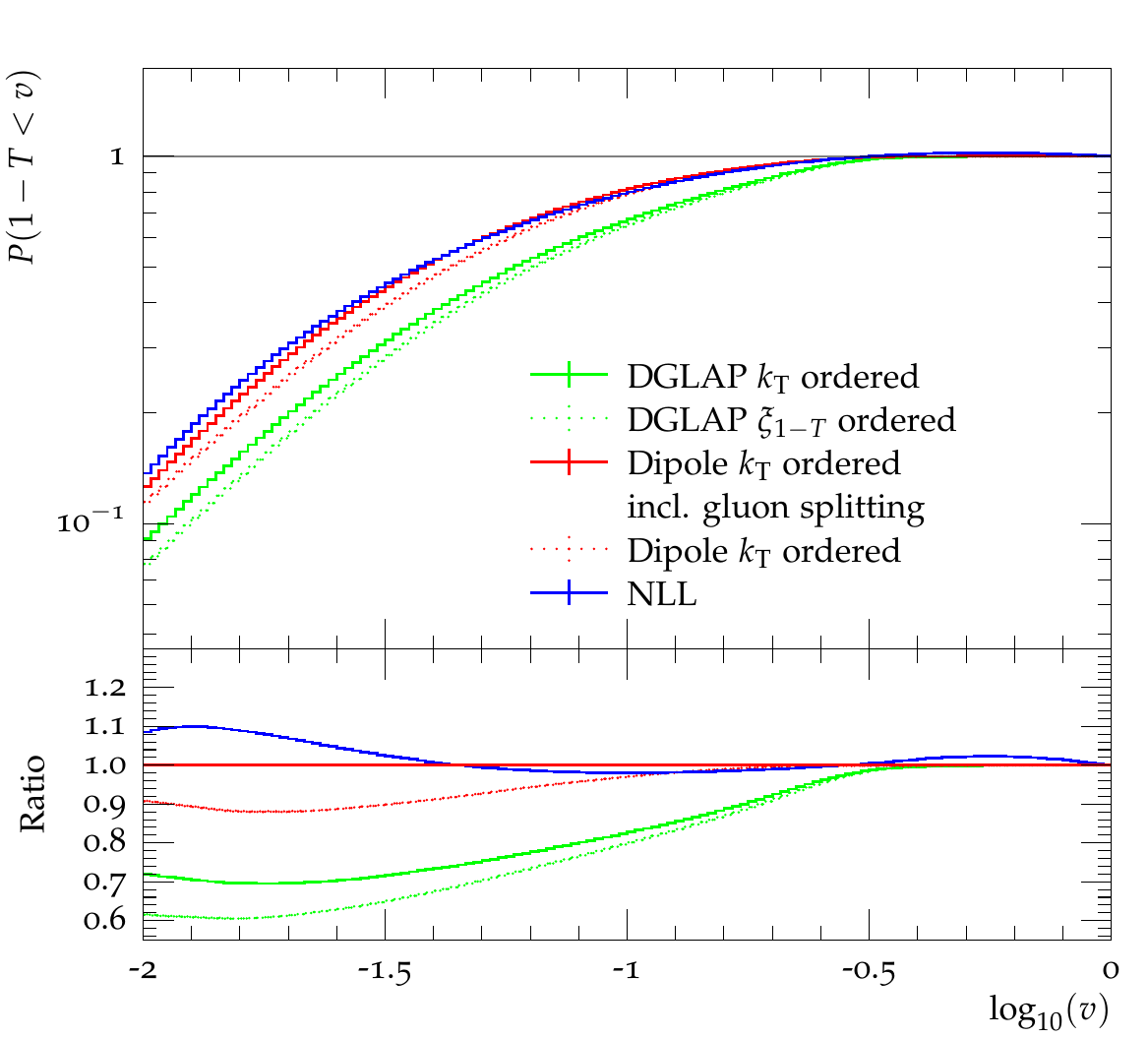}}
  \subfigure[~BKS$_{1/2}$]{\includegraphics[width=.3\textwidth]{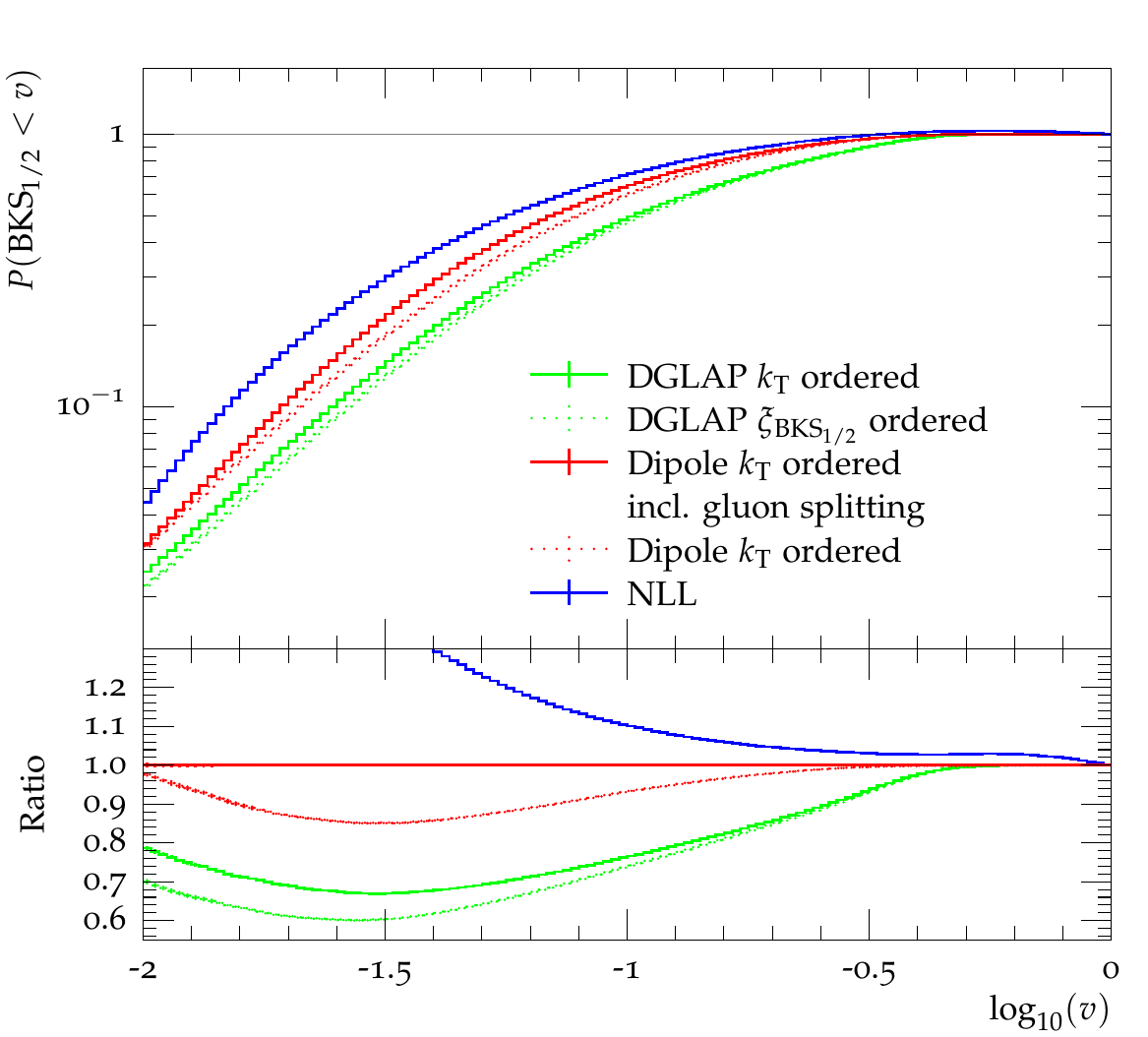}}
  \subfigure[~FC$_{1}$]{\includegraphics[width=.3\textwidth]{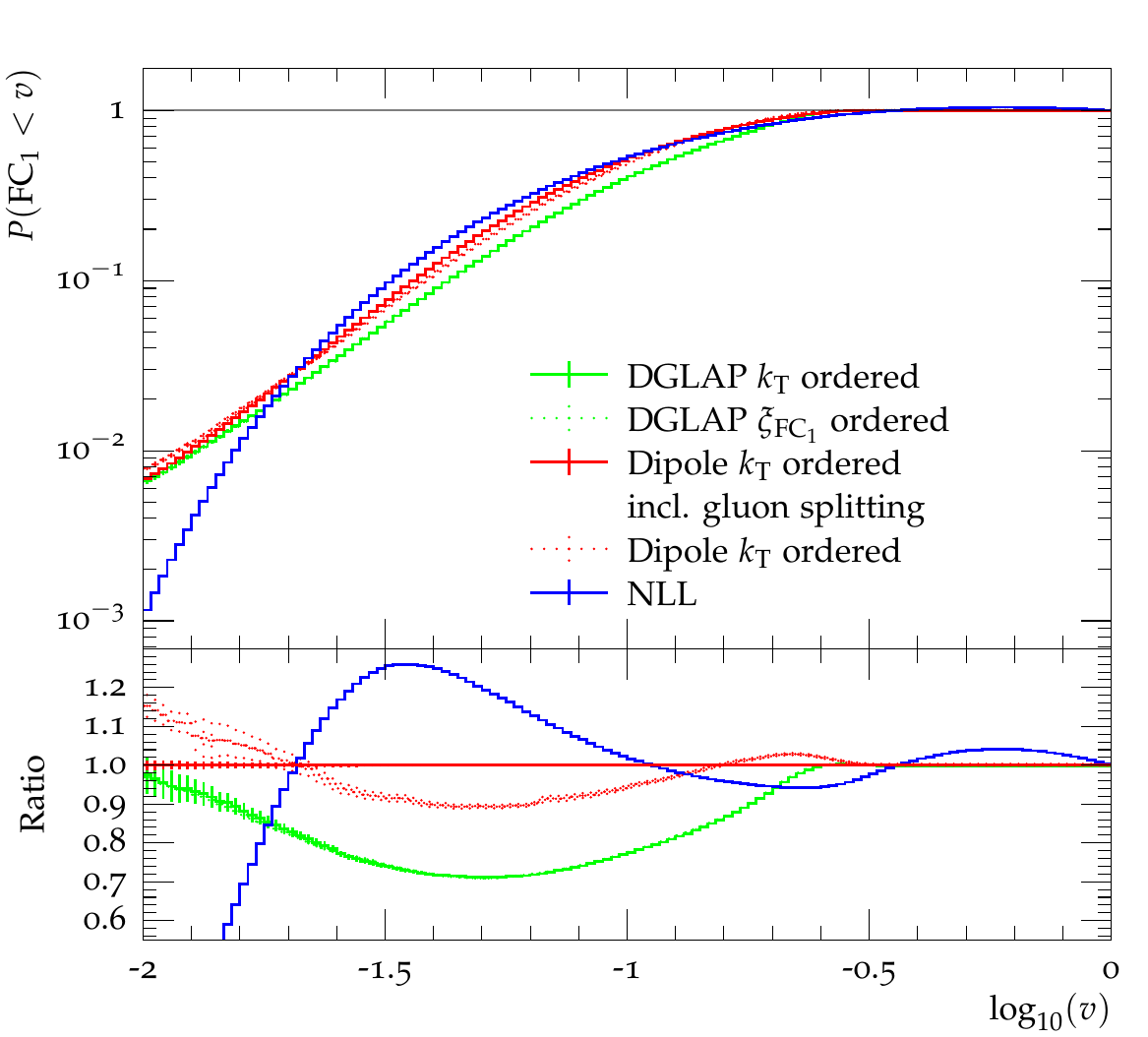}}
  \caption{Comparison between plain DGLAP parton shower ordered in $\xi$ to DGLAP
    parton shower ordered in dipole-$k_T$ and dipole shower with and without gluon splitting.}
  \label{fig:CompareToDipole} 
\end{figure}

Figure~\ref{fig:CompareToDipole} shows a comparison between results from
the dipole-like parton shower in its default configuration (including gluon splitting)
and from a modified version, tailored to match the settings of the parton shower used
in Sec.~\ref{sec:analysis}, Fig.~\ref{fig:FullResult}. It is
interesting to observe that the dipole-shower prediction lies between the parton-shower
result and the analytic result for all observables, and in the case of thrust agrees
very well with the analytic prediction. In the measurable range at LEP energies, the
predictions for FC$_1$ also agree fairly well between the dipole-shower and the
analytic result.

\begin{figure}
\centering
\includegraphics[scale=0.75]{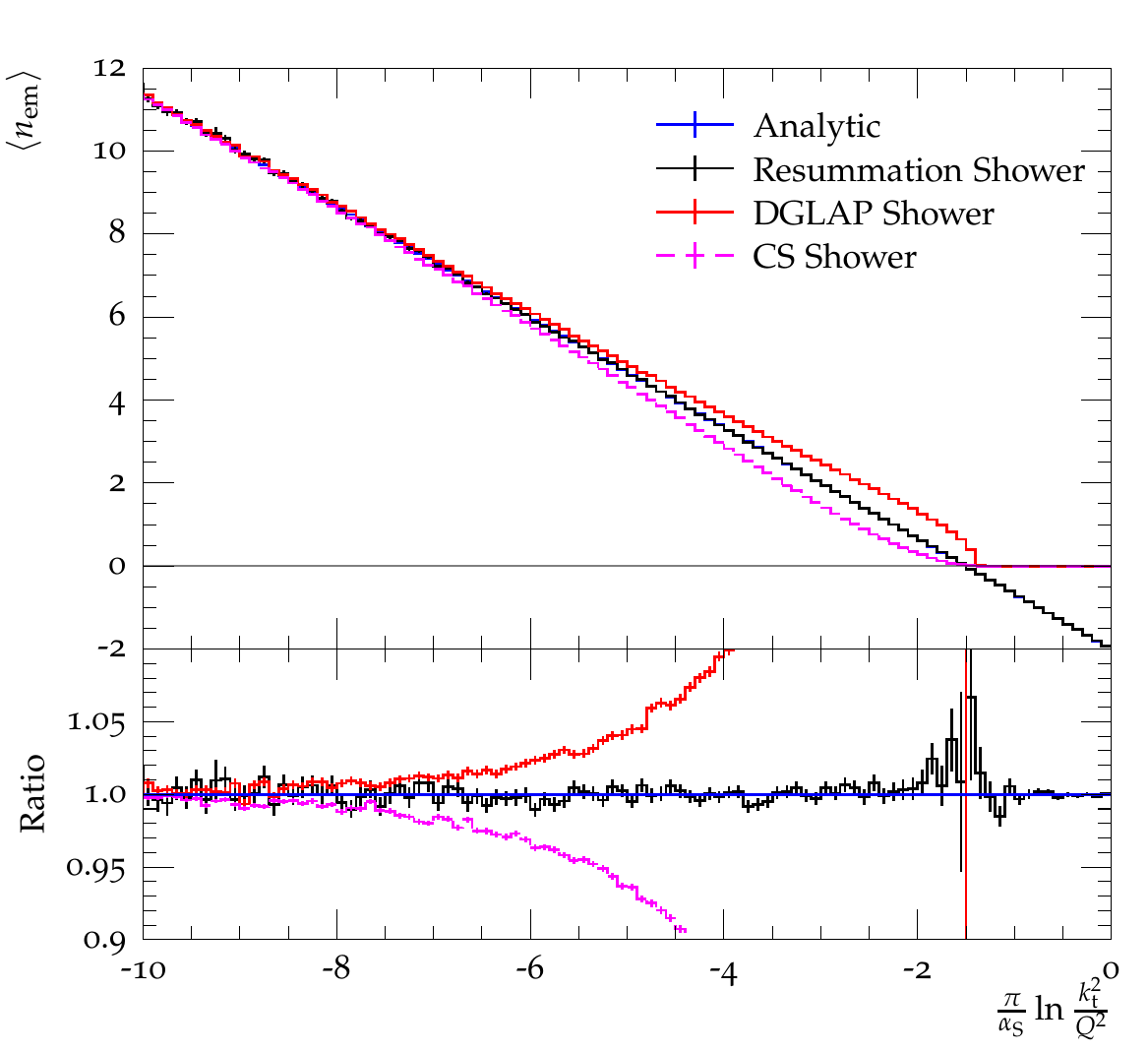}
\caption{Comparison of analytic and parton-shower predictions for the emission
  probabilities in Eqs.~\eqref{eq:single_emission_ps} and ~\eqref{eq:single_emission_nll}.
  The plot shows the average number of emissions per bin as a proxy observable~\cite{Lonnblad:2012hz}.}
\label{fig:R_check}
\end{figure}

Figure~\ref{fig:R_check} displays a cross-check on the logarithmic terms
implemented by the dipole shower as compared to the parton shower and the analytic
result. We extract $R(k_T/Q)$ for a fixed value of the strong coupling,
$\alpha_s=0.118$, using the technique described in~\cite{Lonnblad:2012hz}.
The slope of the distribution corresponds to the leading logarithm, while
the offset of the analytic result corresponds to the next-to-leading logarithm.
Any parton- or dipole-shower prediction must approach the analytic result as 
$k_T\to 0$, which is verified by the convergence of the predictions at small $k_T$.

\section{Conclusions}
\label{sec:conclusions}
We have performed a detailed comparison between pure NLL resummation and parton
showers for additive observables in $e^+e^-$ annihilation to hadrons. We have
isolated their differences, which can broadly be classified as related to
probability or momentum conservation. While a different treatment of these effects
leads to formally subleading corrections on the resummed prediction, it can have
a numerically sizable impact (20\% or more) in the region where experimental measurements
are performed. Similar effects can reasonably be expected to arise in other
observables, as well as in processes with hadronic initial states and with
a more complicated color structure at the Born level. When comparing analytic
resummation to parton showers it should be kept in mind that such differences may
exist, in which case they should be taken into account as a systematic uncertainty.
We have shown in a simple scenario that the differences can be assessed quantitatively
by casting analytic resummation into a Markovian Monte-Carlo simulation
and introducing momentum and probability conservation. Conversely, parton showers
can be modified to violate momentum and probability conservation to reproduce
pure NLL resummation. From the practical point of view this approach is disfavored,
as it leads to numerically inefficient Monte-Carlo algorithms.

\begin{acknowledgments}
  \noindent
  We thank Thomas Gehrmann, Simone Marzani, Stefan Prestel
  and Steffen Schumann for their comments on the manuscript.
  This work was supported by the US Department of Energy
  under contract DE--AC02--76SF00515, by the Deutscher Akademischer
  Austauschdienst (DAAD), and by the German Research Foundation (DFG)
  under grant No.\ SI 2009/1-1.
\end{acknowledgments}

\appendix
\section{Gluon Radiators}
\label{sec:gluon_radiators}
Although we only deal with radiation off quark lines in this study, we argue
that the same conclusions hold for radiating gluons. The basic reasoning is
that the Sudakov form factor, Eq.~\eqref{eq:sudakov} is an alternative form
of the equations in~\cite{Hoche:2017iem} that holds for leading-order DGLAP
splitting functions due to their symmetries. If we use the correct form
of the Sudakov factor, then we can extend the lower integration boundary
for $z$ to zero without encountering a singularity, and we obtain the correct
collinear anomalous dimensions. The detailed argument is as follows.

While the DGLAP equations are schematically identical for initial and final state,
their implementation in parton-shower programs usually differs between the two,
owing to the fact that Monte-Carlo simulations are inclusive over final states.
The evolution equations for the fragmentation functions $D_a(x,Q^2)$
for parton of type $a$ to fragment into a hadron read
\begin{equation}\label{eq:pdf_evolution}
  \frac{{\rm d}\,xD_{a}(x,t)}{{\rm d}\ln t}=
  \sum_{b=q,g}\int_0^1{\rm d}\tau\int_0^1{\rm d} z\,\frac{\alpha_s}{2\pi}
  \big[zP_{ab}(z)\big]_+\,\tau D_{b}(\tau,t)\,\delta(x-\tau z)\;,
\end{equation}
where the $P_{ab}$ are the unregularized DGLAP evolution kernels, and where
the plus prescription is defined such as to enforce the momentum sum rule:
\begin{equation}\label{eq:sf_regularization}
  \big[zP_{ab}(z)\big]_+=\lim\limits_{\varepsilon\to 0}
  \bigg[zP_{ab}(z)\,\Theta(1-z-\varepsilon)
  -\delta_{ab}\sum_{c\in\{q,g\}}
  \frac{\Theta(z-1+\varepsilon)}{\varepsilon}
  \int_0^{1-\varepsilon}{\rm d}\zeta\,\zeta\,P_{ac}(\zeta)\bigg]\;.
\end{equation}
For finite $\varepsilon$, the endpoint subtraction in Eq.~\eqref{eq:sf_regularization} 
can be interpreted as the approximate virtual plus unresolved real corrections, 
which are included in the parton shower because the Monte-Carlo algorithm
naturally implements a unitarity constraint~\cite{Jadach:2003bu}.
For $0<\varepsilon\ll 1$, Eq.~\eqref{eq:pdf_evolution} changes to
\begin{equation}\label{eq:pdf_evolution_constrained}
  \frac{1}{D_{a}(x,t)}\,\frac{{\rm d} D_{a}(x,t)}{{\rm d}\ln t}=
  -\sum_{c=q,g}\int_0^{1-\varepsilon}{\rm d}\zeta\,\zeta\,\frac{\alpha_s}{2\pi}P_{ac}(\zeta)\,
  +\sum_{b=q,g}\int_x^{1-\varepsilon}\frac{{\rm d} z}{z}\,
  \frac{\alpha_s}{2\pi}\,P_{ab}(z)\,\frac{D_{b}(x/z,t)}{D_{a}(x,t)}\;.
\end{equation}
Using the Sudakov form factor
\begin{equation}\label{eq:sudakov2}
  \Delta_a(t_0,t)=\exp\bigg\{-\int_{t_0}^{t}\frac{{\rm d} \bar{t}}{\bar{t}}
  \sum_{c=q,g} \int_0^{1-\varepsilon}{\rm d}\zeta\,\zeta\,\frac{\alpha_s}{2\pi}P_{ac}(\zeta)\bigg\}
\end{equation}
the generating function for splittings of parton $a$ is defined as
\begin{equation}\label{eq:def_updf}
  \mc{D}_a(x,t,\mu^2)=D_a(x,t)\Delta_a(t,\mu^2)\,.
\end{equation}
Equation~\eqref{eq:pdf_evolution_constrained} can now be written in the simple form
\begin{equation}\label{eq:pdf_evolution_constrained_2}
  \frac{{\rm d}\ln\mc{D}_a(x,t,\mu^2)}{{\rm d}\ln t}
  =\sum_{b=q,g}\int_x^{1-\varepsilon}\frac{{\rm d} z}{z}\,
  \frac{\alpha_s}{2\pi}\,P_{ab}(z)\,\frac{D_{b}(x/z,t)}{D_{a}(x,t)}\;.
\end{equation}
The generalization to an $n$-parton final state, $\vec{a}=\{a_1,\ldots,a_n\}$,
resolved at scale $t$ can be made in terms of fragmenting jet functions,
$\mc{G}$~\cite{Procura:2009vm,Jain:2011xz}. If we define the generating
function for this state as $\mc{F}_{\vec{a}}(\vec{x},t,\mu^2)$, we can
formulate its evolution equation in terms of a sum of the right hand side
of Eq.~\eqref{eq:pdf_evolution_constrained_2}. For unconstrained evolution,
we can use Eq.~\eqref{eq:pdf_evolution_constrained}, to write the differential
decay probability as
\begin{equation}\label{eq:pdf_evolution_constrained_4}
  \begin{split}
    \frac{{\rm d}}{{\rm d}\ln t}\ln\bigg(
    \frac{\mc{F}_{\vec{a}}(\vec{x},t,\mu^2)}{\prod_{j\in\rm FS}\mc{G}_{a_j}(x_j,t)}\bigg)
  =&\sum_{j\in{\rm FS}}\sum_{b=q,g}\int_0^{1-\varepsilon}{\rm d} z\,z\,
  \frac{\alpha_s}{2\pi}\,P_{a_jb}(z)\;.
  \end{split}
\end{equation}
Thus, as highlighted in~\cite{Jadach:2003bu}, it is generally necessary to use
the Sudakov factor, Eq.~\eqref{eq:sudakov2}, in final-state parton shower evolution.
At the leading order, the factor $\zeta$ in Eq.~\eqref{eq:sudakov2} simply replaces
the commonly used symmetry factor for $g\to g$ splitting and it also accounts for the
proper counting of the number of active flavors.\footnote{In this context it is
  interesting to note that the factor $\zeta$ has a convenient physical interpretation:
  it represent the ``tagging'' of the resolved parton, for which the evolution is performed.
  This is apparent when extending the evolution to higher orders~\cite{Hoche:2017iem}.}
However, this reasoning applies
only if the boundaries of the $\zeta$-integration are defined by momentum conservation,
and are therefore symmetric around $\zeta=1/2$. In our analysis we attempt to extend
the lower integration limit to zero, which would generate a spurious singularity
arising from the symmetry of the gluon splitting function. Therefore, the commonly
used technique of implementing the symmetrized gluon splitting function without an
additional factor $\zeta$ cannot be used, and the only correct way to treat the problem
is to work with Eq.~\eqref{eq:sudakov2}.

\section{Analytic results at NLL accuracy}
\label{sec:caesar}
This section summarizes the components of the \Caesar formalism~\cite{Banfi:2004yd}
that are needed for our analysis. The resummed cumulative cross section at NLL is given
in this formalism by $\Sigma_{\rm NLL}(v)=e^{-R_{\rm NLL}(Q^2,v)}\mc{F}(v)$,
cf.\ Eq.~\eqref{eq:SigmaCaesar}. The unregularized branching probability $R(v)$
follows from Eq.~\eqref{eq:single_emission_nll}. It is typically written in terms
of $\lambda=\alpha_s\beta_0 L$, where $L=-\ln v$. One obtains
\begin{equation}\label{eq:R_caesar}
  R(v) = 2C_F\left(r(L)+B_qT\left(\frac{L}{a+b}\right)\right)\;,
\end{equation}
where $r(L)$ is separated into a leading and a sub-leading logarithmic
piece as $r(L)=L r_1(\alpha_sL)+r_2(\alpha_sL)$.
\begin{equation}
  \begin{split}
    r_1(\alpha_sL) =&\; \frac{1}{2\pi\beta_0\lambda b}
    \left((a-2\lambda)\ln\left(1-\frac{2\lambda}{a}\right)
    -(a+b-2\lambda)\ln\left(1-\frac{2\lambda}{a+b}\right)\right)\;,\\
    r_2(\alpha_sL) =&\; \frac{1}{b}\bigg(
    \frac{K}{(2\pi\beta_0)^2}\left((a+b)\ln\left(1-\frac{2\lambda}{a+b}\right)
    -a\ln\left(1-\frac{2\lambda}{a}\right)\right)\\
    &\qquad+\frac{\beta_1}{2\pi\beta_0^3}\bigg(
    \frac{a}{2}\ln^2\left(1-\frac{2\lambda}{a}\right)-\frac{a+b}{2}\ln^2\left(1-\frac{2\lambda}{a+b}\right)\\
    &\qquad\qquad\qquad+a\ln\left(1-\frac{2\lambda}{a}\right)-(a+b)\ln\left(1-\frac{2\lambda}{a+b}\right)\bigg)\bigg)\;.
  \end{split}
\end{equation}
The beta function coefficients and the two-loop cusp anomalous dimension
in the $\overline{\rm MS}$ scheme are given by
\begin{equation}
  \begin{split}
    \beta_0 = &\;\frac{1}{2\pi}\left(\frac{11}{6}C_A-\frac{2}{3}T_Rn_f\right)\;,\\
    \beta_1 = &\;\frac{1}{(2\pi)^2}\left(\frac{17}{6}C_A^2-\left(\frac{5}{3}C_A+C_F\right)T_Rn_f\right)\;,\\
    K =&\; \left(\frac{67}{18}-\frac{\pi^2}{6}\right)C_A-\frac{10}{9}T_Rn_f\;.
  \end{split}
\end{equation}
The sub-leading logarithmic term $T(L)$ is defined as
\begin{equation}
  T(L)=\int_{Q^2e^{-2L}}^{Q^2}\frac{dk_T^2}{k_T^2}\frac{\alpha_s(k_T^2)}{\pi}
  =-\frac{1}{\pi\beta_0}\ln(1-2\lambda)\;.
\end{equation}
The $\mc{F}$-function, Eq.~\eqref{eq:f_function}, for additive observables is given by
\begin{equation}\label{eq:f_additive}
  \mc{F}(v) = \frac{e^{-\gamma_E R'(v)}}{\Gamma(1+R'(v))}\;.
\end{equation}
Since both $T(L)$ and $r_2(L)$ are sub-leading in $L$, we have
\begin{equation}\label{eq:Rp_caesar}
  R'(v) = 2C_F\, r'(L)\;,
  \qquad\text{where}\qquad
  r'(L) = r_1'(L) = \frac{1}{b}\left(T\left(\frac{L}{a}\right)-T\left(\frac{L}{a+b}\right)\right)\;.
\end{equation}
Using Eq.~\eqref{eq:Rp_caesar} it can be verified that the combination
of $z_{<v,\rm soft}^{\rm max}$ and $\mu_{<v,\rm soft}^2$ listed in
Tab.~\ref{tab:Parameters} generates the correct value of $r'(L)$,
and therefore the correct value of the $\mc{F}$-function.

\section{Definition of observables}
\label{sec:observables}
This appendix summarizes the definitions of observables used in our study
and lists their parametrizations in terms of the coefficients $a$ and $b$
in Eq.~\eqref{eq:V_approx}. Note that $\vec{q}_i$ stands for any momentum
in the event, no matter if this momentum is hard or soft.

The thrust observable for arbitrary $e^+e^-$ events is defined
as \cite{Farhi:1977sg}
\begin{equation}\label{eq:thrust_def}
  \tau=1-\max_{\vec{n}}\frac{\sum_i |\vec{q}_i\vec{n}|}{\sum_i|\vec{q}_i|}\;.
\end{equation}
The maximization procedure defines a unit vector, $\vec{n}_T$,
which is referred to as the thrust axis. In the 2-jet limit,
Eq.~\eqref{eq:thrust_def} can be written as
\begin{equation}
  \tau=\min_{\vec{n}}\frac{\sum_i |\vec{q}_i|(1-|\cos\theta_i|)}{\sum_i |\vec{q}_i|}\;,
\end{equation}
where $\theta_i$ are the angles of the momenta with respect to $\vec{n}_T$.
The coefficients in Eq.~\eqref{eq:V_approx} are given by $a=b=1$~\cite{Banfi:2004yd}.

The BKS observable is defined as \cite{Berger:2002ig,Berger:2003iw}
\begin{equation}\label{eq:bks_def}
  {\rm BKS}_x=\frac{\sum_i E_i|\sin\theta_i|^x(1-|\cos\theta_i|)^{1-x}}{\sum_i |\vec{q}_i|}\;,
\end{equation}
where $\theta_i$ are again the angles of the momenta with respect to the thrust axis.
For this study we set $x=1/2$, which implies $a=1$ and $b=1/2$~\cite{Banfi:2004yd}.

The fractional energy correlation is defined as \cite{Banfi:2004yd}
\begin{equation}\label{eq:fc_def}
  {\rm FC}_x=\sum_{i\neq j}\frac{E_i E_j|\sin\theta_{ij}|^x(1-|\cos\theta_{ij}|)^{1-x}}{(\sum_i E_i)^2}
  \Theta\big((\vec{q}_i\vec{n}_T)(\vec{q}_j\vec{n}_T)\big)\;,
\end{equation}
where $\vec{n}_T$ is the thrust axis.
For this study we set $x=1$, which implies $a=1$ and $b=0$~\cite{Banfi:2004yd}.

\clearpage
\bibliography{journal}
\end{document}